\documentclass[12pt,showpacs,nofootinbib]{revtex4-1}
\usepackage{amsmath,amssymb}
\usepackage{verbatim,graphicx}

\newcommand\BR{\mathbb{R}}
\newcommand\BZ{\mathbb{Z}}
\newcommand{\BC}{\mathbb {C}}

\def\Tr{\textrm{Tr}}
\newcommand{\lp}{\left (}
\newcommand{\rp}{\right )}
\newcommand{\lb}{\left [}
\newcommand{\rb}{\right ]}

\newcommand{\lr}{\left .}
\newcommand{\rr}{\right .}

\newcommand{\dsp}[1]{$\displaystyle #1$}
\newcommand{\beq}{\begin{equation}}
\newcommand{\beqs}{\begin{equation*}}
\newcommand{\eeq}{\end{equation}}
\newcommand{\eeqs}{\end{equation*}}

\newcommand{\mybox}[1]{
\begin{equation*}
\framebox{\dsp{ #1 }}
\end{equation*}
}

\begin{document}
\setlength{\unitlength}{1mm}
\title{Matrix embeddings on flat $\BR^3$ and the geometry of membranes}

\author{David Berenstein$ ^{\dagger \ddagger}$ and Eric Dzienkowski $^\dagger$\\
$^\dagger$ {\em Department of Physics, University of California at Santa Barbara, CA 93106}\\
$^\ddagger$ {\em Kavli Institute for Theoretical Physics, Santa Barbara, CA 93106 }
}

\begin{abstract} We show that given three hermitian matrices, what one could call a fuzzy representation of a membrane, there is a well defined procedure to define a set of oriented Riemann surfaces embedded in $\BR^3$ using an index function defined for points in $\BR^3$ that is constructed from the three matrices and the point. The set of surfaces is covariant under rotations, dilatations and translation operations on $\BR^3$, it 
is additive on direct sums and the orientation of the surfaces is reversed by complex conjugation of the matrices. The index we build is closely related to the Hanany-Witten effect. We also show that the surfaces carry information of a line bundle with connection on them.
 We discuss applications of these ideas to the study of holographic matrix models and black hole dynamics.
\end{abstract}


\maketitle

\section{Introduction }
\label{S:Introduction}

The discovery of D-branes \cite{DLP} introduced a huge class of new geometric objects in string theory. 
It was quickly realized that the coordinate positions of these geometric objects are matrices, and hence that their positions can become smeared in the same way that typical wavefunctions in quantum mechanics 
do not have a well defined position and momentum due to the uncertainty principle. This fuzzyness of the D-brane position occurs when the matrices that describe the positions of D-branes do not commute. 
When the said matrices commute, the positions of the D-branes can be identified with the eigenvalues of the matrices themselves.

A second route to obtaining noncommutative coordinates for branes arises from the lightcone quantization of the membrane \cite{Hoppe}. Indeed, just the lightcone description in the classical theory itself leads to such a prescription. In that case the coordinates that don't commute describe the internal coordinates of the membrane itself.  The idea of Goldstone and Hoppe is that in the lightcone quantization, the supermembrane acquires a non-degenerate Poisson bracket on its spatial worldvolume. These Poisson brackets on the membrane coordinates are then approximated by commutators of finite matrices, so the matrices become the internal coordinates of the membrane itself. This is a UV truncation in the degrees of freedom, so it regularizes the membrane theory on the lightcone.   Indeed, the supermembrane version of this construction \cite{de Wit:1988ig} was one of the pieces of evidence that was given in the construction of the BFSS matrix model describing the full M-theory on the lightcone quantization \cite{BFSS}. The geometric object we call the membrane is supposed to appear in the limit where the size of the matrices goes to infinity, with small commutators, but it is unclear if a geometric object that is a membrane can be defined for finite matrices or not, and whether it is a sharp geometric object or a very fuzzy object.

All of the considerations above occur at the classical level. There is no need to invoke quantum mechanics to have these noncommutative geometric effects occur. We can then ask the following question: given a collection of matrices that don't commute, is it possible to construct a set of surfaces associated to them that would represent the membrane worldvolume geometry (or multiple membranes) embedded in flat space without taking an infinite size matrix limit?

We will answer this question in the affirmative for the case where we are given three hermitian matrices $X,Y,Z$ that generically do not commute. These would then represent an embedding of a membrane in $\BR^3$. We will also show that this easily generalizes to embeddings on a plane wave in the lightcone where the plane wave has a transverse $\BR^3$ set of coordinates. Our  construction produces a collection of closed oriented surfaces embedded in $\BR^3$ . Indeed, we will show that these membranes not only carry an orientation, but that we can also deduce that they carry vector bundles on them and hence behave like D2-branes.
We will also show that the topology of the membranes is continuous when $X,Y,Z$ are deformed, so topology transitions are not instantaneous and require going through singular geometries. This implies that the associated brane charge is conserved.

The main motivation to understand this problem in detail arises from the study of simulations of black hole formation in matrix models performed in \cite{ABT}. The data obtained there at the end of the evolution is exactly of the sort above: a collection  of (somewhat random) matrices that do not commute. Although one can try to find the brane positions by diagonalizing each matrix, the end result is not amenable to easy visualization in higher dimensions. After all, what are we supposed to do with the noncommutative information?  The approach in \cite{AHHS} is to find some approximate locations for D0 branes that minimizes the non-diagonal matrix elements given this choice of basis and then throws the off-diagonal information away. Then it uses those D0 branes as a proxy for the geometric object. This is a very nice idea.  However, this type of description does not describe the orientation of the extended membranes nor can it be used to determine the topology of the brane configuration except in the large matrix limit. This is especially hard if one is near a topology change. It also seems to indicate that the result is very fuzzy and the topology of the branes is in the end given by the topology of a set of points.

Our approach to this problem is very different. We start from the BFSS \cite{BFSS} and BMN matrix models \cite{BMN} and to simplify matters in the discussion, we orbifold the problem sufficiently so that in the end we can deal with a reduced model where only three matrices are required, so instead of starting with a system with $16$ supersymmetries, we go to a system with $4$ supersymmetries. We do this by taking a model which is the dimensional reduction of $\BZ_k$ supersymmetric orbifolds  in four dimensions that are chiral. The model  we need is then obtained by taking a quiver where only fractional branes on one of the nodes of the quiver are present, this is,  we concentrate on the reduction of pure $N=1$ SYM reduced to matrices (there are 3 of them that are dynamical, which we call $X,Y,Z$). To explore the geometry we then add a fractional brane probe in one of the nodes of the quiver diagram that intersect the node where our matrices are located. The main idea is to ask what the D-brane probe sees given a generic set of coordinates as given by the $X,Y,Z$ matrices. Within the dynamics of the matrix model orbifold, there are either bosons connecting the probe brane to the configuration, or fermions. The interesting degrees of freedom to define the geometry end up being the fermions. Indeed, one can show that the spectrum of fermions connecting the probe to the fuzzy object can be obtained by diagonalizing a simple hermitian matrix obtained from the $X,Y,Z$ and the coordinates of the probe. The matrix can be thought of as an effective Hamiltonian and it is given by
\begin{equation}
H_{eff} = (X-x) \sigma^x +(Y-y) \sigma^y+(Z-z) \sigma^z \label{eq:main}
\end{equation}
A technical point is that the eigenvalues of $H_{eff}$ are not all  positive nor all negative. 
In second quantizing the fermions the positive eigenvalues are associated to raising operators, and the negative eigenvalues to lowering operators. The conjugate modes come from the anti-chiral fermions. The absolute value of the eigenvalues then serves as the mass for the fermions. This mass can be thought of as the length of a fermionic string connecting the probe to the configuration, so it gives a notion of distance, and the minimal eigenvalue is the shortest distance to the configuration.

The  important technical point of this paper is that the eigenvalues of $H_{eff}$ can cross zero and this depends on the position of the probe. The geometric surface locus of the matrix configuration is described exactly by the locations where one of the eigenvalues of $H_{eff}$ vanishes. Counting the number of positive eigenvalues of $H_{eff}$ versus the negative eigenvalues can indicate the number of such crossings of zero, and this can be used to define an index $I(x,y,z)_{X,Y,Z}$. The plane $\BR^3$ represented by the coordinates $x,y,z$ is then colored by the index of the location. The index is locally constant and can only change if one of the eigenvalues crosses zero. Any path connecting two points with different index must have zero crossings: it is impossible to avoid them by taking a clever path. This indicates that the surfaces obtained this way are closed. At any crossing, we also get an orientation: from higher index to lower index, so the surfaces are oriented.

A second point that is worth mentioning is that at the zero crossing a raising operator becomes a lowering operator, or viceversa. Thus if one follows a vacuum of the fermion degrees of freedom on one side of the brane continuously to the other side we get an anomalous creation of fermionic strings. This is a generalization of the Hanany-Witten effect \cite{HW} and it represents anomalous creation of branes by branes (for more details on the relations to topology and anomaly inflow of this effect see \cite{BDG}).

The paper is organized as follows. In section \ref{sec:BFSS} we review the BFSS and BMN matrix models and some of their orbifolds. This is used to find an effective Hamiltonian for chiral fermions in the presence of a configuration of three matrices plus a probe, which was already written in equation \eqref{eq:main}. Next, in section \ref{sec:index} we show how the number of positive versus negative eigenvalues of the effective hamiltonian can be used to define an index function given a position of a probe. The index vanishes when the probe is at infinity. The locus where the index changes defines a collection of surfaces. We show various properties of the index. In section \ref{sec:fuzzy} we show that fuzzy spheres give configurations where on some loci the index is non-zero. The associated surfaces are spheres. We also show how to construct torus embeddings in $\BR^3$. Finally, we show that the surfaces carry the information of a line bundle on them. This
makes the behave like D-branes. In section \ref{sec:link} we show how to generalize the index to a linking number between two such configurations. The linking number ends up counting the number of strings that are created by trying to separate the two configurations in a generalization of the Hanany-Witten effect. Equally, the index counts the number of strings that are created when bringing the configurations together from infinity. We show that in a special case of fuzzy spheres for the BMN model, that the data on when the fermion zero modes appear provides additional evidence  that the surfaces carry a line bundle on them and behave like D2-branes.  We apply these ideas to study matrices obtained by numerical studies of the BMN matrix model in section \ref{sec:bh}. We give applications to understand the polarization of black holes into membranes and we also give applications where the Hanany-Witten effect can stop probe D-branes: we show that as $N$ increases, the Hanany-Witten effect strings end up storing more energy parametrically in $N$  than the probe, so they are enough to show that the black hole will stop the probe inside it and it will not come out at the other side. Finally, in section \ref{sec:conc} we conclude. We also have appendices   where some of our more elaborate computations and conventions are described.

\section{The BFSS and BMN matrix models, and their orbifolds}\label{sec:BFSS}

The BFSS matrix model \cite{BFSS} is simply the dimensional reduction of $9+1$ SYM to $0+1$ dimensions. It is a gauged matrix quantum mechanics of $9$ adjoint matrices $\phi^i$ with
a gauge symmetry by similarity transformations. 
An amazing aspect of this dynamical theory is that it is a description of M-theory in the discrete lightcone quantization and in the limit $N\to \infty$ and in an appropriate double scaling limit it gives a quantization of M-theory in flat space in a lightcone quantization. The action is given by
\begin{align}
S_{BFSS} &= \int dt\, \Tr\lb \sum_{j=1}^9 \frac{1}{2(2R)}(D_0\phi^j)^2 + \frac{i}{2}\Psi^\dagger D_0\Psi + \frac{(2R)}{4}\sum_{j,k=1}^9[\phi^j,\phi^k]^2 \rr \nonumber \\
&\quad \lr + \sum_{j=1}^9 \frac{1}{2}(2R)(\Psi^\dagger\gamma^j[\phi^j,\Psi]) \rb
\end{align}
As written, the action depends on a parameter $R$. This parameter can be eliminated by a rescaling of the variables and time and it can be replaced by $\hbar$. Thus the BFSS matrix model itself has no intrinsic scale at the classical level. Indeed, one can check that even $\hbar$ can be removed because the action has a classical scaling symmetry. 

The simplest version of the model is for $1\times 1$ matrices. In this case, the classical configurations are given by a point in $\BR^9$ with a velocity, and the fermions just add degeneracy to these states. This degeneracy gives the correct degrees of freedom for a graviton supermultiplet in 11D \cite{BFSS}. The $\BR^9$ describes the transverse  directions to the lightcone, and the rank of the matrices is the amount of lightcone momentum. Such a configuration is a D0-brane.
Ground states in the classical theory in  general correspond to configurations of commuting matrices, where $N$ such D0-branes are located on $\BR^9$. 

One of the important things about the BFSS matrix model is that it is capable of describing extended objects. Indeed, one can describe D2-branes and higher order D-branes from these configurations. These are easy to see in the infinite N limit \cite{Banks:1996nn}, as central charges can be activated in the supersymmetry algebra that encode such objects of infinite extent. Such solutions lead to effective noncommutative field theories. A modern introduction to the geometric interpretation of these developments can be found in \cite{Steinacker:2010rh}.

One can also check that matrix configurations source the various supergravity fields at long distances, and that the couplings to weakly curved backgrounds give us a way to compute the currents and the multipoles with respect to the brane charges of the configurations. This was very systematically developed in the works of Taylor and collaborators \cite{Taylor}. A review of the BFSS matrix model where all of this is very clearly addressed is in \cite{Taylorreview}. 

Finite matrix configurations can also behave like extended D-branes.
The simplest example of such configurations are fuzzy spheres 
\cite{Kabat:1997im}, where three of the matrices are made proportional to angular momentum matrices. These have been studied extensively. An important question to ask is if these geometries survive at finite $N$, or if they are only well defined strictly when $N\to \infty$.
We will show in this paper that precise geometries can be described even for finite $N$. However, since in the BFSS matrix model in principle one can also describe all other D-branes in type IIA string theory, a random configuration of matrices would be too complicated: it would probably encode somewhat random extended D-branes of type IIA theory. It would be nice if we could reduce the problem to just studying surfaces in three dimensions, where only three of the $\phi$ matrices matter, and the other six are eliminated somehow.  

A very simple way of doing this is by realizing that the BFSS matrix model can also be thought of as the dimensional reduction of $N=4$ SYM in four dimensions down to $0+1$ dimensions. If we manage to reduce the supersymmetry from $N=4$ SYM to just $N=1$ SYM, then instead of having $9$ matrices $\phi^i$, we would get only three matrices, those that arise from the dimensional reduction of the gauge field connection. In that situation, the D0 branes would be confined to an $\BR^3$, rather than an $\BR^9$, and we might expect that we can only describe D2 branes, as any higher dimensional even brane would have too high a dimension to fit in three dimensions. 

A  simple way to achieve this truncation and to keep a full geometric interpretation of the system in terms of string theory is to take a supersymmetric orbifold $\BC^3/\BZ_k$. These are described by quiver theories which can be constructed by the techniques developed by Douglas and Moore \cite{DM}. What matters for us is that we can end up with such theories where only the dynamics of $N=1$ SYM matter. 

 This would be the theory of $N$ identical fractional branes at the orbifold singularity. A  simple explanation of how those field theories can be built and studied is found in \cite{KaS}. The geometric interpretation in terms of fractional branes and intersection theory of those objects can be found in \cite{DGM}.

 For simplicity we can choose a $\BZ_k$ action that gives rise to chiral theories where between any two nodes in the quiver there is at most one chiral field connecting them: this way if we add a probe for a different fractional brane, we can get a single chiral multiplet worth of fields connecting the probe to the configuration. The particular example of an orbifold we can choose is given by acting on $\BC^3$ given by coordinates $\alpha_1,\alpha_2, \alpha_3$ and acting with the $\BZ_k$ defined by the identifications  $\alpha_1\to \omega \alpha_1$, $\alpha_2\to \omega^2 \alpha_2$, $\alpha_3\to \omega^{-3} \alpha_3$, and $\omega= \exp(2\pi i/k)$ is a primitive root of unity. Many other orbifolds will have similar properties and the precise details of the orbifold are not important at this stage.  All that we need can  be visualized by a simple subquiver diagram
 \begin{equation}
U (N)\  \bullet \longrightarrow \bullet \ U(1)\ \hbox{ Probe}
 \end{equation}
where the arrow indicates a single chiral multiplet.

 The advantage of having a single chiral field is that the fermions are represented by a two component Weyl spinor, and the $\gamma$ matrices appearing in the BFSS matrix model reduce to the four dimensional gamma matrices for such spinors: those are just the Pauli matrices. The details of the reductions are shown in the appendix. The other advantage of having chiral fields is that they carry anomalies in four dimensions, thus it is natural to assume that they might encode a lot of topological information even in the reduction to $0+1$ dimensions. 

Upon such a reduction, we end up with fermion terms where we only involve four dimensional $\gamma$-matrices. Indeed, if we reduce to a single chiral multiplet, then we can think of the $\gamma$-matrices themselves as Pauli matrices. The effective action is then
\begin{align}
S_{orb} &= \int dt\, \Tr\lb \sum_{j=1}^3 \frac{1}{2(2R)}(D_0\phi^j)^2 + \frac{i}{2}\Psi^\dagger D_0\Psi + \frac{(2R)}{4}\sum_{j,k=1}^3[\phi^j,\phi^k]^2 \rr \nonumber \\
&\quad \lr + \sum_{j=1}^3 \frac{1}{2}(2R)(\Psi^\dagger\sigma^j[\phi^j,\Psi]) \rb
\end{align}
where if $\psi$ is chiral, then $\psi^\dagger$ is antichiral. Again, $R$ is meaningless as it can be redefined away, and the classical symmetries of $S_{orb}$ have the same properties as those for $S_{BFSS}$. The action above is a shorthand: it is the same action of the BFSS matrix model, but the matrices are restricted by the orbifold conditions  \cite{DM} \footnote{ In practice this can be done keeping the form of the action fixed and adding information about the matrix restrictions by using a crossed product algebra \cite{BL} . This will produce a set of orthorgonal projectors for each node of the quiver, and the commutation relations with these projectors will recover all the information of the quiver diagram. For example the traces of the projectors will recover the rank of the gauge groups on each node .}.

The new advantage is that now we only have to deal with three hermitian matrices $\phi^{1,2,3}$ rather than $9$. Also, the Pauli matrices are easier to handle than the 9 dimensional gamma matrices.

The question is then if given $\phi^{1,2,3}$ hermitian matrices, can we associate a collection of D2-branes in a specific geometric configuration in $\BR^3$ to it? To the extent that we can, we can then uplift any such intuition to 9 dimensions and understand better how membrane geometries arise in the BFSS matrix model.

Another useful matrix model to consider is the BMN matrix model \cite{BMN}. That model describes $M$ theory on a plane wave in the discrete lightcone quantization. Its action is given by
\begin{align}
S &= S_{BFSS} + S_{mass} \\
S_{mass} &= \int dt\, \Tr\lb \frac{1}{2(2R)}\lp -\lb \frac{\mu}{3}\rb^2\sum_{j=1}^3(\phi^j)^2 - \lb\frac{\mu}{6}\rb^2\sum_{j=4}^9(\phi^j)^2\rp - \frac{i}{2}\lp\frac{\mu}{4}\rp\Psi^\dagger\gamma_{123}\Psi \rr \nonumber \\
&\quad \lr - \frac{\mu}{3}i\sum_{j,k,l=1}^3\epsilon_{jkl}\phi^j\phi^k\phi^l\rb
\end{align}
which is a mass deformation of the BFSS matrix model.
Again, if we look at $1\times 1$ matrices, the configuration space is $\BR^9$, but there are no flat directions: there is a quadratic potential in the Hamiltonian. This is as it is supposed to be: it just reflects the fact that there is a gravitational potential in the plane wave geometry.  This model also can be obtained by a dimensional reduction of $N=4$ SYM to $0+1$ dimensions. We need an $SU(2)$ invariant reduction on a sphere \cite{KKP}. Again, dealing with full 9-dimensional matrices is not very intuitive, so we can play the same orbifold trick on $\phi^{4\dots 9}$ to get rid of those matrices. Again, we can get rid of $R$ and we can choose units so that $\mu=3$, but then we are not free to rescale $\hbar$ to be whatever we want any longer. Thus the BMN matrix model does have a parameter $\hbar$, also when we orbifold. One can go towards the classical regime, and again, if we give three matrices $\phi^{1,2,3}$ we can ask: is there a way to associate a geometric D2-brane to such a configuration? Indeed, once we take the BMN matrix model at finite $N$, the ground states are made of collections of concentric fuzzy spheres. These are such that the $\phi$ themselves are angular momentum matrices with canonical normalization.

The answer both here and in the BFSS matrix model will be yes: one gets an associated set of surfaces in both cases. The surfaces will be slightly different in the BFSS versus the BMN matrix model. The new ingredient in the BMN matrix model is that the fermions get a contribution proportional to $\gamma^{1,2,3}\propto \sigma^1 \sigma^2\sigma^3\propto 1$ in their mass. This reflects the fact that in the 11 dimensional maximally supersymmetric plane wave background there is a non trivial background flux. Such a contribution changes slightly the shape of the branes that one associates to the configurations and this is essentially due to the Myers effect \cite{Myers}: branes are polarized in the presence of RR backgrounds. In particular a D0 brane becomes polarized into a sphere.

The essence of this article is to look in detail at the fermion degrees of freedom to understand the geometry of branes. Our technique is that given three $X$ matrices (these are identified with $\phi^{1,2,3}$ for one of these models and will be called $\vec X$ collectively, or $X,Y,Z$ if we want to name them individually), we will ask: what would a probe point like D0-brane see? We ask the question from the point of view of the fermion degrees of freedom that connect it to the configuration we  are studying. This is encoded in the dynamics of the fermions themsleves. The problem reduces to studying an effective Hamiltonian given by
\begin{equation}
H_{eff} \simeq (\vec X -\vec \lambda) \cdot \vec\sigma \quad + \left( \frac 34\right)
\end{equation}
The last term is for the BMN model and it describes the additional contribution to the effective Hamiltonian from flux. The value comes from a choice of orientation of the flux and how it relates to the chirality of the fermions (the sign choices when we take $\gamma^{1,2,3}$). 
Solving for $H_{eff}$ gives us the energies of the fermions that connect the object to the configuration. In the full dynamics, the wave function solutions for $H_{eff}$  are  second-quantized, as the fermionic objects $\psi$ become operators when we turn on quantum mechanics. This is important for the physical interpretation of the membranes.

\section{The index: adding a D0-brane probe}\label{sec:index}

As we have described previously, the geometry in the BFSS and BMN matrix models is encoded by matrices of dimension one. In this section we will work exclusively with the BFSS matrix model type of dynamics. We will only add some passing remarks at the end. To understand how a generic object of the model looks geometrically (a general matrix configuration), we can ask the question by adding a point-like probe. This is, we want to extend the size of the matrices by one, by taking a direct sum with a zero-brane probe. This just means that we put the $N\times N$ matrix configuration and embed it into an $(N+1)\times (N+1)$ matrix in the upper left corner, we add the eigenvalues in the rightmost bottom corner and add zeros everywhere else. This is, we have a new auxiliary configuration where
\begin{equation}
\tilde X = \begin{pmatrix} X&0\\
0&x \end{pmatrix}, \quad \tilde Y = \begin{pmatrix} Y&0\\
0&y \end{pmatrix}, \quad \tilde Z = \begin{pmatrix} Z&0\\
0&z \end{pmatrix} \label{eq:probe1}
\end{equation}
In our problem, the matrices $X,Y,Z$ share their properties with the BFSS matrix model: they are three hermitian matrices.

The essence of the geometric characterization of the general matrix will then be encoded in the observations of a spectator brane. The spectator brane will only be allowed to ask questions related to the dynamics of the matrix models themselves and in particular of the degrees of freedom that connect the extra eigenvalue to the matrix configuration. 

In general there are two classes of modes that connect the extra eigenvalue to the configuration: bosonic degrees of freedom and fermionic degrees of freedom. We will restrict ourselves to the fermionic degrees of freedom. The questions we will ask depends on the position of the extra eigenvalue probe.

When we look at the fermions, we decompose them as follows
\begin{equation}
\tilde \psi=\begin{pmatrix} 0& \psi\\
0&0\end{pmatrix}\label{eq:probe2}
\end{equation}
where our goal now is to ask what are the energies associated to the modes $\tilde\psi$. Notice that this picked a very particular component of the fermions and not the other. This can be justified completely in orbifold models, as we discussed previously, but orbifolds are not really required to make this argument. All we need is the subquiver diagram that enforces the restrictions of the matrices defined by equations \eqref{eq:probe1} and \eqref{eq:probe2}. To do this carefully, we are choosing the probe to be a different fractional brane than the matrices $X$ represent. The chirality of these modes indicates that if the branes were four dimensional fractional branes, then they would intersect for sure if we think of  fractional branes as higher dimensional branes wrapped on collapsed cycles.   The intersection properties of the fractional branes represent the intersection properties of the collapsed cycles \cite{DGM}. Notice also that we did not put fermions in the bottom leftmost corner: this is our chirality assumption for the arrow.

The obvious question to ask first, is if there is a definition of distance from the eigenvalue probe to the matrix configuration. The way to ask that question is to look at the spectrum of fermions connecting the probe eigenvalue to the matrix configuration. The eigenvalue probe is located at $x,y,z$ and the three matrices $X,Y,Z$ are three $N\times N$ hermitian matrices. Given our three matrices $X,Y,Z$, this is described by the following effective Hamiltonian:
\begin{equation}
H_{eff} = ( X-x 1_{N}) \sigma_x+(Y-y 1_N) \sigma_y+ (Z-z1_N)\sigma_z\label{eq:hef}
\end{equation}
where we are being pedantic in stating that $x$ is multiplying the identity matrix of $N\times N$ matrices. We will omit this in the future. The structure of how the Pauli matrices appear for chiral multiplets is derived in the appendix \ref{fermiondecomposition}. 

This is the Hamiltonian of the fermionic degrees of freedom connecting the probe brane to the rest of the configuration. The origin of this Hamiltonian is seen from the term in the full Hamiltonian given by
\begin{equation}
 \Tr (\psi^*\gamma^i [X_i, \psi])
\end{equation}
when evaluated in the configuration $\tilde X, \tilde Y, \tilde Z$. The Hamiltonian above describes the mass term for the off diagonal modes of the fermion $\psi$ that are charged under the gauge group of the extra eigenvalue probe. 
The gamma matrices in three dimensions are given by the Pauli matrices, whereas the dependence on $x$ and $X$, etc, comes from direct evaluation of the commutators. 

We can think of this as a Hamiltonian in a tensor product space $Hilb_{big}= Hilb_N \otimes Hilb_{\uparrow\downarrow}$ of an $N-$ dimensional Hilbert space times a spin one half object ( a single q-bit).
$H_{eff}$ is covariant under unitary transformations of $Hilb_N$. This is, we have that under $U\in Aut(Hilb_N)$, we can consider this inducing an automorphism of 
$Hilb_{big}$ by $U \otimes 1$. The automorphism takes $X\to U XU^{-1}$, $Y\to U YU^{-1}$, $Z\to U ZU^{-1}$ and $H_{eff}\to ( U \otimes 1) H_{eff}( U^{-1} \otimes 1) $ which shows that the spectrum of $H_{eff}$ is invariant under such rotations. This is inherited from the gauge transformations of the original matrix model. What is important is that the spectrum of $H_{eff}$ is gauge invariant.

As is usual in string theory, the off diagonal modes connecting a subconfiguration to another are considered to be strings, once they are quantized. The typical energy of 
a string of length $\ell$ is given by $\alpha' \ell$ where $\alpha'$ denotes the string tension. Hence, in our effective Hamiltonian, we can denote the distance from 
the probe brane located at $(x,y,z)$ to the configuration by the eigenvalues of the effective Hamiltonian $H_{eff}$. The reason to look at fermions is that fermionic Hamiltonians do not have tachyons. Thus technically all energies are positive and thus the notion of distance is positive. This is also true for string states: open string fermions in the NSR superstring appear in the Ramond sector for open strings. The zero point energy of the fields cancels between bosons and fermions on the worldsheet (they have the same boundary conditions) and the only contribution to the energy of the string is from the classical stretching between the ends of the strings.  

The eigenvalues of $H_{eff}$ themselves can be positive or negative, so we interpret the positive eigenvalues as the frequencies of creation operators, and the negative eigenvalues as frequencies of lowering operators once we second-quantize. The absolute value of the spectrum of $H_{eff}$ is then the list of distances from the probe brane to the object when interpreted as strings. Obviously, if we have more than one distance, the object with respect to which we are measuring distances should be considered to be an extended object. The minimal eigenvalue of the spectrum thus obtained should give us the minimal distance to the extended configuration.

Notice that the Hamiltonian $H_{eff}$ is covariant under rotations and translations. This is inherited from the symmetries of the original BFSS Lagrangian. More importantly, the Hamiltonian is also covariant under rescalings (if we rescale $X,Y,Z$ and the coordinates $x,y,z$ by the same common factor, the entries of the matrix rescale with the same power and thus the eigenvalues scale).

Let us solve the problem of the spectrum first in the asymptotic regime, where lets say $(x, y, z) \to \infty$ along a determined direction in $\BR^3$ keeping the $X,Y,Z$ matrices fixed. By convenience, we can use rotation invariance to take $z\to \infty$ keeping $x,y$ equal to zero.

Then we have that 
\begin{equation}
H_{eff} = -z \sigma_z + ( Z \sigma_z + X\sigma_x+Y\sigma_y)
\end{equation}
We can compute the eigenvalues of $H_{eff}$ by considering it as a perturbation theory of $H_{eff}\simeq  - z\sigma_z$. The eigenvalues of this matrix are degenerate. There are $N$ eigenvalues of values $+z$ and $N$ eigenvalues of value $-z$. These are very large. Since the spectrum is degenerate, to first order we need to resolve
the splitting among the degenerate subset. This is done by looking at the perturbation terms in $H_{eff}$ that commute with $\sigma_z$. The term that does that is $Z \sigma_z$ itself. So the transformation that diagonalizes $H_{eff}$ along the two degenerate subsets is the same transformation that diagonalizes $Z$. 

We find that the leading order spectrum is given by
\begin{equation}
Eig(H_{eff}) = \pm (z-\lambda^z_i)+ O(1/z)\label{eq:perturbationtheory}
\end{equation}
 where $\lambda^z_i$ are the eigenvalues of $Z$. The extra corrections of order $1/z$ are from perturbation theory: they result from `energy denominators' and involve the components of $X,Y$. 

We find the familiar theme that the eigenvalues of the matrices $X,Y,Z$ describe the positions of objects (distances) as viewed from infinity. Since the eigenvalues are continuous functions of the matrices, we find that the notion of distance by taking the minimum eigenvalue is a continuous function of the position. 

We are now interested 
in asking what happens when we are at distance zero from a configuration. 

This can happen in two ways: an eigenvalue of $H_{eff}$ crosses zero, or the eigenvalues just graces zero and keeps its sign. A really interesting question is whether the spectrum of $H_{eff}$ always has paired eigenvalues: if eigenvalues cross zero, this is not so. If eigenvalues are always paired, then every time one eigenvalue reaches zero
from positive values, then another one reaches zero from negative values. We will define an index that counts possible crossings of zero from infinity. At infinity, the spectrum of $H_{eff}$ is paired into positive and negative eigenvalues and to first order in perturbation theory they are equal to each other up to sign, obviously this implies that they both have the same number of eigenvalues. If an eigenvalue goes from positive to negative, the number of positive eigenvalues decreases by one, and the number of negative eigenvalues increases by one. Similarly in the other direction. We want the index to be zero at infinity and to change by one by each crossing. The definition of the index is given by
\begin{equation}
I( (x,y,z))_{X,Y,Z}  = \frac{n_+ - n_-}{2}
\end{equation}
where $n_+$ the number of  positive eigenvalues of $H_{eff}$, and $n_-$ is the dimension of the space of negative eigenvalues of $H_{eff}$. The 
index is a locally constant function (after all, eigenvalues of matrices are continuous functions of the entries)  that can only change values at locations where $H_{eff}$ has null eigenvalues. 
If for a configuration we have that $I(x,y,z)\neq 0$, we know that on any path connecting $x,y,z$ to infinity there are crossings of zero and thus the location $(x,y,z)$ is surrounded by the noncommutative object characterized by $X,Y,Z$.

Such an index was defined in \cite{HL} for the position $x,y,z=0$. It was called a Bott index. In their formulation, they were dealing with approximations to a sphere, where $X^2+Y^2+Z^2\simeq 1$ and the introduction of Pauli matrices was an auxiliary construction in mathematics. The matrices $X,Y,Z$ represented observables in a quantum system where only finitely many states are allowed and hence position observables become finite matrices. They were also restricted to have small commutators. The 
operator $H_{eff}$ in that case would square to something that was very closed to the identity, so all eigenvalues of $H_{eff}$ would need to be very close to $\pm1$.  Counting positive and negative eigenvalues is an invariant under small deformations that prevent the eigenvalues from getting too far from $\pm1$.
The index as interpreted in that case was an obstruction to localizing the states on a sphere (making $X,Y,Z$ strictly commute), by demanding that $||X^2+Y^2+Z^2-r^2|| <\delta$ by deformations of $X,Y,Z$ that keep this property and a bound on their commutators is implemented. The spectrum of the operator $\vec X \cdot \vec \sigma$ is also used in numerical  studies of noncommutative field theories ( see \cite{DelgadilloBlando:2012xg} for a recent example), and one can also use the operator $\vec X\cdot \sigma$ to define a fuzzy sphere by studying a single matrix model of $2N\times 2N$ matrices with a constraint \cite{Steinacker:2003sd}.

 In the case we have described here the index is dictated by the dynamics of fermionic degrees of freedom on D-branes. There are also no restrictions on the size of commutators. These ideas can be extended further to higher dimensions and matrices with various restrictions following the ideas in \cite{HL2}. Such a generalization is beyond the scope of the present paper.

Here are basic properties of the Index function (some of these  already appear in the work  \cite{HL}):
\begin{enumerate}

\item The index is an integer. At infinity the index vanishes ( as we computed already). The index changes by $\pm 1$ if a single eigenvalue crosses zero. It changes by integers if many eigenvalues cross zero.

\item Orientation: the index defines a collection of oriented closed surfaces. The surfaces are the locus where the index changes value. The orientation is defined by
going from larger values to smaller values of the index (this includes the sign, thus $-1>-2$ etc). The surface set itself is obtained from the zero locus of a polynomial in $(x,y,z)$ obtained by taking determinants. These surfaces will be called membranes or D-branes interchangeably. 

\item Additive property. Given two configurations $X_1,Y_1,Z_1$ and $X_2, Y_2,Z_2$, we can consider a new configuration given by taking direct sums
$X_3= X_1\oplus X_2$, $Y_3= Y_1\oplus Y_2$, $Z_3=Z_1\oplus Z_2$. The index is additive under such constructions
\begin{equation}
I((x,y,z))_{X_3,Y_3,Z_3}= I((x,y,z))_{X_1,Y_1,Z_1}+I((x,y,z))_{X_2,Y_2,Z_2}
\end{equation}
and the set of surfaces with orientation is also additive under this operation.

\item Orientation reversal. This states that we can reverse the orientation of a surface without affecting its shape. This is done by considering the complex conjugate to the matrices $X,Y,Z$. In equations, we have that
 \begin{equation}
  I( (x,y,z))_{X,Y,Z}=- I( (x,y,z))_{X^*,Y^*,Z^*}
\end{equation}
This property is less obvious. A proof is as follows:  A matrix and its transpose have the same eigenvalues. Thus $(\vec X-\vec x)\otimes \vec \sigma$ has the same eigenvalues as  $(\vec X-\vec x)^T\otimes \vec \sigma^T$. Now, $\vec X^T= \vec X^*$, so we can substitute. However, for Pauli matrices we have that $\vec \sigma^T \simeq -\vec\sigma$ after a unitary transformation in the spin one half subspace. Thus. we have that the eigenvalues of $(\vec X-\vec x)\otimes \vec \sigma$ are equal to the eigenvalues of $(\vec X^*-\vec x)\otimes (-\vec \sigma)$. This is, the matrix $(\vec X^*-\vec x)\otimes \vec \sigma$ has the same eigenvalues as $(\vec X-\vec x)\otimes \vec \sigma$ but with signs changed. This exchanges $n_+$ and $n_-$ and reverses the index.

\item If $X,Y,Z$ are real, then $I((x,y,z))=0$ everywhere. This is a corollary of the orientation reversal property. Obviously for such configurations we have that 
$  I( (x,y,z))_{X,Y,Z}=- I( (x,y,z))_{X^*,Y^*,Z^*}=- I( (x,y,z))_{X,Y,Z}$. From which the result follows. This in particular holds for collections of zero branes: direct sums of one dimensional problems.

\item The index is covariant under rotations, translations and dilatations of the system. This follows from the similar properties that $H_{eff}$ has.

\item The index is not trivial: there are matrix configurations $(X,Y,Z)$ for which $I((x,y,z))_{X,Y,Z}\neq 0$. We will explore these in the next section.
 
\end{enumerate}

If we instead work with the BMN matrix model we get an effective Hamiltonian given by
\begin{equation}
H_{eff} = -z \sigma_z + ( Z \sigma_z + X\sigma_x+Y\sigma_y)+\frac 34 \sigma^x\sigma^y\sigma^z
\end{equation}
The extra term causes trouble with scaling the surfaces, and with being able to change the sign of the eigenvalues by complex conjugation. This way various of the properties above are broken. For example the change of orientation does not happen automatically, and the corresponding Index does not behave as nicely. We still get translation and rotation covariance. The index still vanishes when a probe is at infinity, but one can check that even for $1\times 1$ matrices, the index changes when the probe which is used to define the index is on top of the 0-brane that described the configuration. This is the Myers effect in action \cite{Myers}. Indeed, as far as fermions are concerned, the presence of a background RR flux changes the Dirac equation, and an example computed by one of the authors of the paper can be found in  \cite{BJL}. In that example the displacement of the location of the fermion zero modes was required in order for configurations to form tori that were BPS. In the present case, the structure of the gamma matrices follows the background flux in the BMN model \cite{BMN}.

On the other hand, in this case many fuzzy spheres are ground states of the system and one expects that these solutions survive as time independent configurations. Also, many of these can be made to oscillate slightly so the membranes can persist indefinitely.

\section{Fuzzy spheres and emergent surfaces}\label{sec:fuzzy}

\subsection{Fuzzy spheres}

Now that we have defined an index, let us consider some special examples of the index computation. We will start with a fuzzy sphere and ask about the index at the center of the sphere. The fuzzy sphere is defined as follows. Let $L^{1,2,3}$ be the angular momentum matrices of the irreducible representation of $SU(2)$ of spin $j$. These satisfy the identities
\begin{equation}
[L^{i},L^j]= i \epsilon^{ijk} L^k
\end{equation}
The maximum eigenvalue of $L^3$ are $\pm j$. 
Consider the following set of 3 matrices built by the following combinations:
\begin{equation}
X= \frac{r}{j} L^1, Y = \frac {r}{j} L^2, Z= \frac{r}{j} L^3
\end{equation}
This is called a fuzzy sphere. The maximum eigenvalue of $Z$ is $|r|$. Thus one could argue that the sphere has radius $|r|$ (as seen from infinity as in our large distance computation in the previous section). 
Notice that $X^2+Y^2 +Z^2 = \frac{j(j+1)}{j^2} r^2$. Thus one could also argue that the radius of the sphere is given by
$\tilde r= \sqrt{(1+\frac 1j)}| r|$. These two become identical in the large $j$ limit, but at finite $j$ there is some discrepancy. However, it is natural to believe that there is a well defined surface near this radius that surrounds the origin and that is our candidate for a locus where an eigenvalue changes sign.

Let us prove this assertion by computing the index in the center of the configuration, at $x=y=z=0$. The effective Hamiltonian we have to deal with is then given by
\begin{equation}
H_{eff} = \frac r j \vec L \cdot \vec \sigma
\end{equation} 
This is the same type of problem that  shows up in  spin-orbit coupling in the hydrogen atom. The important thing is that this is spherically symmetric, so it makes sense to decompose the Hilbert space $Hilb_{big}$ into irreducible representations of $SU(2)$. The big  Hilbert space is given by
 \begin{equation}
 Hilb_{big}\simeq (j) \otimes (\frac 12)  \simeq (j+ \frac 12) \oplus (j- 1/2)
 \end{equation}
and it decomposes into two irreducible representations of $SU(2)$. For each of them, we have a common eigenvalue of $H_{eff}$. Moreover, $H_{eff}$ is traceless. This can be proved in general because the Pauli matrices themselves are traceless. Thus, the two possible eigenvalues of $H_{eff}$ have the opposite sign. One is positive, and the other is negative. This depends on the sign of $r$. Let us choose the sign of $r$ so that $n_+>n_-$. The number of eigenvalues of the bigger representation of $SU(2)$ is
$n_+=2j+2$, while those of the smaller representation are $n_-=2j$. These are the dimensions of the two irreducible representations of $SU(2)$ appearing in the tensor product.
We obtain that 
\begin{equation}
I((0,0,0))_{\hbox{Fuzzy Sphere}} = \frac{n_+-n_-}{2} = 1
\end{equation}
We already knew that the index was an integer, and that the typical change should be by $\pm 1$. Here we find an explicit example where the index changed  by one somewhere between the origin and infinity. Because of spherical symmetry, the index changes value at a fixed sphere radius. A direct computation carried in the appendix shows that the radius at which it happens is given exactly by $|r|$. We thus find that the radius is governed by the maximum eigenvalue, rather than by the value of $X^2+Y^2 +Z^2$. Indeed, if we use the definition of distance from the origin that is obtained from the spectrum of $H_{eff}$ we find that the distance is equal to $|r|$. Indeed, with the spectral definition of distance we used, we find that the distance from any point in space to the sphere is the one that is obtained by elementary geometry. 

Obviously, we can also set up direct sums of concentric fuzzy sphere configurations of various radii, so we can get configurations where the index is arbitrarily large. For such configurations the index counts the (minimal) number of sphere layers that need to be crossed to get out of the center. Since the index counts with sign, surfaces (which we call membranes) of different orientations can be present and the index itself represents a lower bound on the number of layers that need to be crossed.

\subsection{From a sphere to a torus}

Here we detail how to make configurations that lead to a fuzzy torus embedded in three dimensions. The idea is to begin with a fuzzy sphere and to deform the matrices in a simple form to go from a sphere to a torus. The basic idea is to follow the construction of the giant torus as described in \cite{NT} (other examples of embeddings of Riemann surfaces in $\BR^3$ can be found in \cite{Arnlind:2006ux}, and in \cite{Shimadaft} one can also find a different example that interpolates between sphere and tori).  In the case of the giant torus, one is supposed to add strings with maximal angular momentum to a sphere until the geometry transitions to a torus. To do this, it is convenient to use matrices defined by 
\begin{equation}
X^{\pm} = X\pm i Y
\end{equation}
and in the other direction
\begin{eqnarray}
X= \frac{X^+ + X^-}{2}\\
Y=\frac{X^+ - X^-}{2i}
\end{eqnarray}
The matrices $X^{\pm}$ in the fuzzy sphere of spin $j$ case are rescaled ladder operators for spherical harmonics. $X^{\pm}$ are adjoints of each other.
In a natural basis for a sphere, we have that
\begin{equation}
X^+_{ba}= r \sqrt{ (j(j+1)-a(a+1)} \delta_{b, a+1}
\end{equation}
The labels $a,b$ go from $j\dots - j$. 

In the matrix $X^+$, if we quantize fluctuations of the fuzzy sphere (see for example \cite{DSJR}), one can check that the different diagonals of the matrix carry different amounts of angular momentum in the $z$ direction. They differ by one unit, and the diagonal where $X^+$ has entries carries no angular momentum in the $z$ direction. When we condense various of these fluctuations, we simply replace them by an expectation value which becomes just a number multiplying the appropriate fuzzy tensor harmonic. Since we are looking to maximize the angular momentum of the fluctuations,  the deformation we seek is given by
\begin{equation}
X^+_{ba}= r \sqrt{ (j(j+1)-a(a+1)} \delta_{b, a+1}+ r  \beta \delta_{b,j} \delta_{a,-j}
\end{equation}
and we then take $X^-= (X^+)^\dagger$. We are using the index convention for the matrices that is associated to the $L_z$ spin of the $SU(2)$ representation of spherical harmonics. The self-adjoint matrices $X$,$Y$ are built from the same linear combinations as above, after the deformation.
The parameter $r$ just rescales the full solution, so we can ignore it. The parameter $\beta$ then controls the geometry. For $\beta=0$ we have a sphere. Indeed, the topology of the sphere is preserved for some values of $\beta$ around zero. We have seen numerically that the topology changes at the precise value $\beta=j$, this is not essential for our discussion.

\begin{figure}[ht]
\includegraphics[scale=0.5]{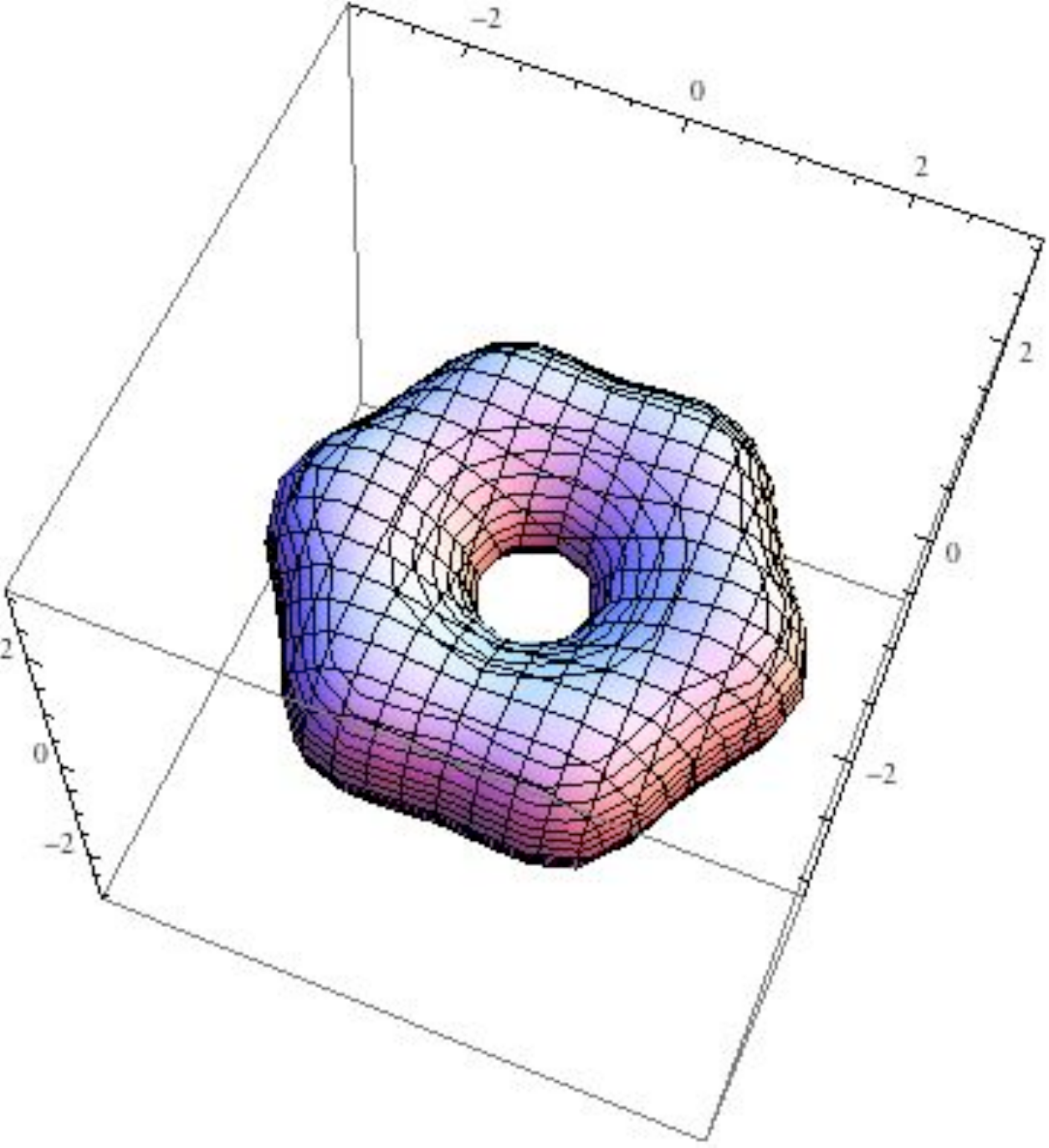}
\caption{A fuzzy torus, for $r=j= 5/2$, $\beta=2.55$. The $\BZ_6$ symmetry is easily visible.}\label{fuzzytorus}
\end{figure}

Another thing to notice is that the presence of $\beta$ breaks the rotational symmetry to $\BZ_{2j+1}$ which is the rank of the matrices. This can be understood from the spin of the excitations around the $z$ axis: it is the unbroken symmetry associated to condensing the state with highest spin along the z axis.  Thus the torus shape is not invariant under full rotations along the $X,Y$ plane. The simplest case where the family of surfaces we get seems to contain a torus is for $4\times 4$ matrices.
A figure for the case of $6\times 6$ matrices is presented in figure \ref{fuzzytorus}. 

We should also notice that in our case it is obvious we have a torus. In other setups, to argue for the genus of the surface is more involved, using an approximation to Morse theory on the surface \cite{Shimada:2003ks}, and the result is inherently more fuzzy, or explicitly requires taking a limit of large matrices \cite{AHH}. One can also obtain more standard fuzzy tori as zero energy configurations in higher dimensions by studying beta deformed matrix models \cite{Shimada:2008xy}. 

\subsection{D2-branes}

The main characteristic of D-branes is that they carry a connection on their worldvolume. This is the familiar statement that the open string sector has a massless spin one particle on the D-brane worldvolume. Here, we want to show that the geometry that we deduced for these surfaces carries the information of a line bundle on it.

The idea is rather simple. The surfaces are defined by the vanishing of an eigenvalue of $H_{eff}$, as calculated for equation \eqref{eq:hef}. Obviously, for a single zero eigenvalue there is a corresponding eigenvector. Let us call it $\psi_0 (x,y,z)$.

The eigenvector $\psi_0$, normalized to unity,  is well defined up to a $U(1)$ phase. This is the familiar symmetry for states in a Hilbert space in quantum mechanics: a global phase for the full wave function is not measurable, as physical states are rays in the Hilbert space.

Now, for each position on the surface there is such an eigenvector. This changes continuously when we vary the position along the surface $(x,y,z)$, as the eigenvectors are also smooth functions of the matrix entries. One can easily understand this fact by the fact that the eigenvectors  can be calculated using perturbation theory in quantum mechanics.

One can construct a bundle from these $\psi_0(x,y,z)$. One defines sections of the bundle by functions multiplying $\psi_0(x,y,z)$. Because the phase of $\psi$ is ambiguous, we have to choose a phase by patches on the surface, and between patches there are transformation rules for $\psi_0$. 

One can also define a connection on the patches. This is done by the familiar Berry phase, defined by
\begin{equation}
 v^\mu A_\mu = -i v^\mu \psi^*_0(x,y,z) \partial_\mu \psi_0(x,y,z) =-iv^\mu\langle \psi_0| \partial_\mu|\psi_0\rangle
\end{equation}
where $v^\mu$ is a tangent vector to the surface. 
Obviously, this defines the connection of a line bundle on the worldsheet. Thus, at least in principle, the membrane behaves exactly like we would expect a D2-brane to behave. At this stage, it is not clear the Berry connection that one would compute this way is just the connection that open strings feel, or if this is further twisted by the tangent bundle on the surfaces that were defined as we prescribed.

The full exploration of the curvature on these bundles and the precise connection to D-branes is beyond the scope of the present paper. We will show later that there is further evidence for physical states feeling a connection on the membrane worldsheet when we intersect two of these objects.

\section{A linking number}\label{sec:link}

Now that we have defined a geometric object for a collection of $3$ hermitian matrices, we can do something more. We can take two such objects and ask how they are related to each other. Indeed, in the matrix model setups, each of them would be a configuration of branes, so the spectrum of strings stretching between them becomes interesting from a dynamical point of view. One can define a linking number that for zero branes at a position $\vec x$ reduces to the index we defined in previous sections.

The idea is to take the matrices $X_1, Y_1, Z_1$ of rank $r_1$ and $X_2, Y_2, Z_2$ of rank $r_2$ and define a matrix analog the Hamiltonian $H_{eff}(x,y,z)_{X_1,Y_1,Z_1}$,
where we replace $(x,y,z)$ by hermitian matrices $(X_2,Y_2,Z_2)$. If the matrices commute with one another, we want the $H_{eff}$ operator to give us an operator that acts as $H_{eff}$ on the direct sum over
the eigenvalues of $X_2,Y_2,Z_2$. One easily sees that the following Hamiltonian does that:
\begin{equation}
H^{(1)}_{eff}(\vec X_1, \vec X_2) =( X_1\otimes 1_{r_2}-1_{r_1} \otimes X_2)\otimes\sigma_x +( y\leftrightarrow x) +(z\leftrightarrow x)
\end{equation}
Notice that once $H^{(1)}$ is defined this way, it does not matter anymore that the $X_2, Y_2, Z_2$ matrices commute with each other.

Then, the definition of our linking number is given by
\begin{equation}
L^{(1)} [(X_1, Y_1, Z_1), (X_2,Y_2,Z_2)] = \frac{n^1_+-n^1_-}{2}
\end{equation}
It's easy to prove that $L^{(1)}$ is antisymmetric in the entries. This is because tensor product spaces $A\otimes B$ are equivalent to $B\otimes A$ as Hilbert spaces.
If we think of these spaces in tensor notation, the equivalence is a reordering of the indices. The Hamiltonian $H_{eff}^{(1)}$ then changes sign (more precisely, $H_{eff}^{(1)}(\vec X, \vec X')$ is unitarily equivalent to $-H_{eff}^{(1)} ( \vec X', \vec X))$ when we exchange the triples $\vec X_1$ and  $\vec X_2$.

There is a second linking number that one can define, by changing a brane by an antibrane, this is, reversing orientation:
\begin{equation}
H^{(2)}_{eff} =( X_1\otimes 1_{r_2}-1_{r_1} \otimes X^*_2)\otimes\sigma_x +( y\leftrightarrow x) +(z\leftrightarrow x)
\end{equation}
Again, if $X_2, Y_2, Z_2$ commute with each other and are diagonal, we can not distinguish $H^{(2)}_{eff}$ from $H^{(1)}_{eff}$. But if the matrices do not commute with each other, we can. The definition of the second linking number is 
\begin{equation}
L^{(2)}[ (X_1, Y_1, Z_1), (X_2,Y_2,Z_2)] = \frac{n^2_+-n^2_-}{2}
\end{equation}
This is symmetric in the exchange of $(X_1,Y_1,Z_1)$ and $(X_2,Y_2,Z_2)$. This uses the antisymmetry of $L^{(1)}$ combined with the change in sign of the index 
upon complex conjugation discussed in previous sections. It turns out that when considering the dynamics of fermions as given in the BFSS matrix model, it is the spectrum of $H_{eff}^{(2)}$ that controls the physics \cite{TR}. This is because the matrix multiplication rules on commutators translate to needing to take the transpose of the matrices $X_2,Y_2,Z_2$, which is equivalent to using their complex conjugates. This is also equivalent to saying that the fermions transform as a fundamental under one set of branes and an antifundamental with respect to the other set of branes.

Also notice that if we move one of the objects and take them to infinity (by adding multiples of the identity matrix), then at infinity both of the linking numbers are zero. Also, if we shrink one object until it is point-like  (by making $X_2, Y_2, Z_2$ proportional to the identity matrix, with coefficients $x_2, y_2, z_2$), then the linking number is $r_2$ times the index $I(x_2, y_2,z_2)_{X_1,Y_1,Z_1}$. 

Also, one can use these same Hamiltonians $H^{(1)}_{eff}$ and $H^{(2)}_{eff}$ to define a spectral distance between two such configurations, again by taking the eigenvalues closest to zero and taking absolute values. For infinite membranes touching each other in the IKKT matrix model one finds zero modes \cite{Chatzistavrakidis:2011gs}. The effective Hamiltonian for fermions in that case takes a similar form to the BFSS matrix model. This is just as expected from the mode spectrum of brane intersections at angles \cite{Berkooz:1996km}. When the intersections are extended and compact, the low lying modes at the intersection need to be quantized carefully and zero modes are not guaranteed. One would expect that the spectral distance then gives an upper bound for a geometric distance between the brane configurations.

We will now give an application of the linking numbers. We will show that the linking numbers actually take into account that the surfaces that are defined by previous sections actually carry a connection for a line bundle on them that couples to physical states. This provides further evidence that the surfaces are actually behaving as D2-branes. This is easiest to check from the calculations in the appendix. 

The idea is as follows: take two fuzzy spheres and displace them relative to each other. For simplicity, we have them normalized so that the radius is equal to $j$ and $j'$, the spin of the corresponding representations of $SU(2)$. This is natural in the BMN model for ground states.  Let the displacement between the fuzzy sphere centers be characterized by $b$. Because of the high amount of symmetry, one can actually compute the index analytically and follow the crossings of zero of the fermion eigenvalues in a lot of detail. If the displacement is $b$, along the $z$ axis, and the fermions are decomposed into fuzzy spherical harmonics with respect to both fuzzy spheres, one finds that the eigenvalues cross zero sequentially when $b= j-j'+ \ell$, where $\ell$ is an integer between $0$ and $2j'$ inclusive. This means the index of the configuration where the big fuzzy sphere surrounds the smallest is exactly equal to $2j'+1$, which is the dimension of the representation of the spin $j'$ set of matrices. This is expected: the small fuzzy sphere is made of $2j'+1$ D0 branes, so that when they are all inside the big fuzzy sphere, we expect the index to be $2j'+1$ times the index of the smallest representation.

The first zero mode appears when the spheres touch each other for the first time, at displacement $b=j-j'$. As $b$ advances further, the two fuzzy spheres touch each other along a circle. We expect the lightest fermions to be localized in this circle. So the problem effectively reduces to a one dimensional problem. As can be seen from the results of the appendix, the fermion modes with maximal angular momentum in each $SU(2)$ representation of fuzzy spherical harmonics do not mix with other states, and their frequencies are given exactly by
\begin{equation}
j-j'+\ell- b 
\end{equation}
where $\ell$ is an integer. These states are evenly split in energy, creating effectively a Kaluza-Klein tower of finitely many states (this is very similar to the Kaluza-Klein tower of tachyons between such spheres computed in the BMN model for such crossings in \cite{BT}). Such a Kaluza-Klein tower is an approximation to a quantum field theory for zero mass fermions on a circle (either leftmoving or right-moving depending on if the energy of the mode is positive or negative) in the presence of an holonomy around the circle (this can be translated to quasi-periodic boundary conditions on the fermions if we want to). The fermions can have zero eigenvalues if the holonomy is a multiple of $2\pi$. This can be removed by a large gauge transformation redefining the notion of momentum on said circle. The important thing to notice is that if the corresponding surfaces that are intersecting have the properties of D-branes, in that each carries a connection ${\cal A}_1$, ${\cal A}_2$ , then the fermions that stretch between them feel the connection ${\cal A}_1-{\cal A}_2$. Because of spherical symmetry, this connection can be computed from Gauss law by calculating the area of the sphere that the circle where the fermions lie enclose $\oint {\cal A }_1 ds= \int_{S_1} F_1 da$. The area of a sphere slice is proportional to the height of the slice, hence 
$\int_{S_1} F_1 da\propto A $ which is linear in the vertical height of the slice. 
\begin{figure}[ht]
\includegraphics[scale=0.8]{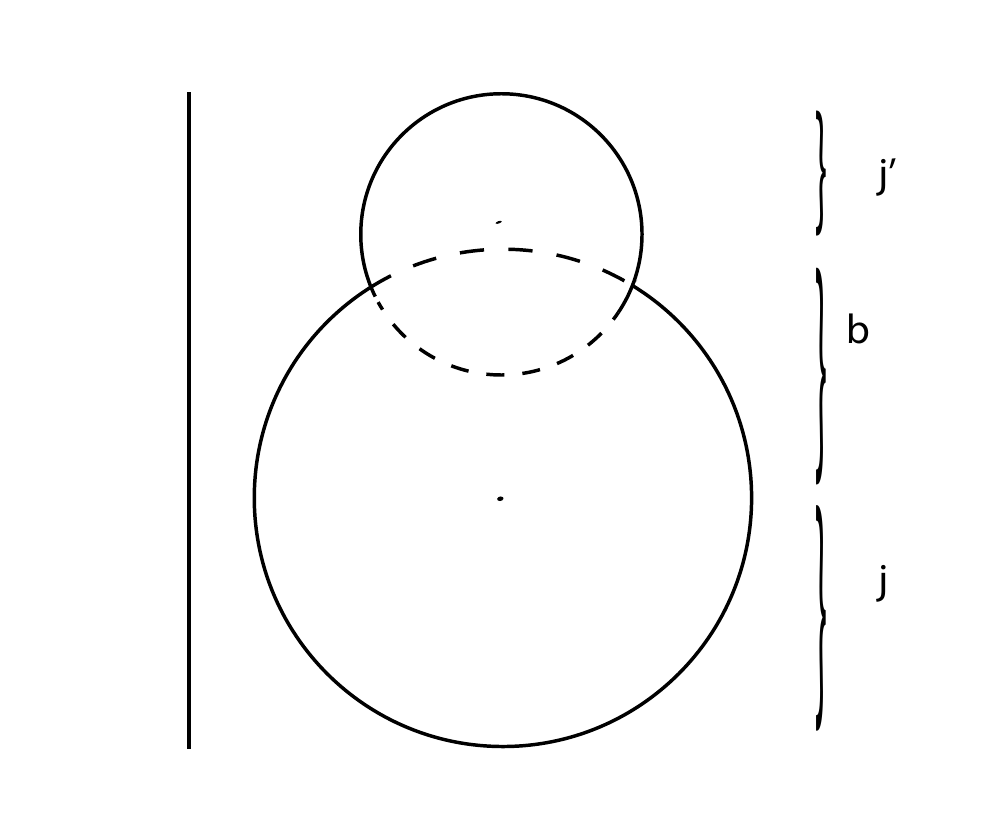}
\caption{Illustration of two intersecting spheres. The net connection seen by the fermions stretching between can be computed by calculating the net flux through the solid lines. }\label{fig:spherestouch}
\end{figure}

  Hence, shifts of $2\pi$ in the holonomy are equally spaced in $b$. Indeed, if the sphere is made of $n$ D0 branes, we expect the total flux through the sphere of this bundle to be equal to $n= 2j+1$. However, we can also expect a curvature correction. If we think of the matrices as describing a lowest Landau level of endpoints on each sphere, in order to have $n$ states we need a monopole flux equal to $n-1$ (this is if the endpoints are treated as monopole spherical harmonics). This extra one is the contribution of the curvature of the sphere. One can check this way that the net flux for this connection through each sphere  is $2j$ and $2j'$ respectively, as opposed to $2j+1$ and $2j'+1$. Thus the net flux through a slice is  proportional to the area, which is $2\pi j (2 j-t)$ where $t$ is the height of the slice. Since the total flux through each sphere is $2j$, and the area is $4\pi j^2$ for each sphere, we get that the flux per unit height
is constant and the same for both spheres. Thus, the flux for each sphere is linear in height with the same coefficient. This can be visualized in figure \ref{fig:spherestouch}.
As seen in the figure, the net flux that we need to compute is the one associated to the surface that has not been dashed in the graphic. This is proportional to $j+j'+b$ as a function of the displacement. We need this number to be an integer in order to get the correct holonomy. We see then that the geometric argument matches the matrix computation. Obviously this is a simplified computation where it so happens that the flux per unit height on each sphere is the same.

Thus, the setup shows that indeed the surfaces we make are compatible with the idea of having a curvature of a line bundle on them for physical states that thread between them. There is another way to think about this that we already discussed: on the locus of positions where an eigenvalue vanishes there is a preferred fermion wavefunction for the hamiltonian $H_{eff}$: this is the zero eigenvector itself. This is only well defined up to a phase. If we want to patch these together to form a vector bundle over the surface, we need a line bundle connection so that this phase ambiguity is resolved on parallel transport. This is the generic case, but we can set it up so that the null eigenspaces are degenerate (thereby giving us multiple branes on top of each other). Thus, in general one will need a bundle connection to resolve these issues. Since the structure that we are analyzing involves the symmetries of a Hilbert space under change of basis, the connection in general will be $U(n)$ valued for $n$ coinciding branes. 

The last thing that we will do in this section is to give a more physical interpretation of these zero modes. The main idea, which we have hinted at already when we defined the index function, is that a crossing by zero represents a raising operator becoming a lowering operator and viceversa (for the particle conjugate). If we follow a ground state continuously past this change, the ground state is defined by $a|0\rangle=0= b |0\rangle$, where $a$ is the lowering operator for particles (the ones with positive frequency), and $b$ is the lowering operator for the antiparticles (the ones associated to negative frequencies in $H_{eff}$). After the crossing by zero, the state that follows by continuity of $|0\rangle$ is not ground state anymore. Instead, one of the lowering operators, let us say $a_\alpha$, becomes a $b^\dagger$ (a raising operator). The state on the other side of the barrier will have a non-zero occupation number for a single fermion in the Hilbert space. This is, on crossing a zero, a fermion is created. This is nothing but the Hanany-Witten effect (and its various generalizations discussed in \cite{HW,BDG}). 

The linking number we defined then encodes the number of strings that are created (with orientation) when separating two objects that are partially inside each other. Or the number of strings that were created on bringing the objects together from infinity when they cross each other. This is done by following a vacuum adiabatically until exactly the point where the transition happens (where there is a degeneracy of vacua), and then following the state that is created after a crossing  and which is not a vacuum any longer adiabatically as well, until further crossings where fermion zero modes determine a degeneracy of a Fock space of fermions at each level. 

Notice that this interpretation in terms of the Hanany-Witten effect explains why the geometry is so sharp. The presence or not of strings connecting the two surfaces is easy to test: we check if we the fermionic ground state is gauge invariant or not. The Hanany-Witten effect has the property that the fermionic ground state is not always gauge invariant, so the presence of the strings is protected by topology. 

\section{Aspects of matrix black holes}\label{sec:bh}

Recently, various simulations have been carried in matrix models to understand various aspects of the dynamics of black holes in holographic setups. The main idea so far has been to compare the numerical simulations in the BFSS matrix model with  black holes as described in \cite{IMSY}. The numerical approach was initiated in the works \cite{CW, AHNT} and a lot of the thermodynamic static properties of the black holes have been matched in the quantum mechanics. The most impressive such agreement is in \cite{HHNT}.
The BFSS matrix model 
has an infinite moduli space of vacua, so the thermal ensemble of these models is not well defined. 
This gives such calculations a systematic error.
To 
have a better setup one wants a matrix theory with a well defined ensemble, and the BMN matrix model fits the bill. Numerical simulations using lattice techniques were carried out in \cite{CVa}. Also, classical simulations of the dynamical evolution of the BMN matrix model have been carried out in \cite{ABT}.

All of these calculations in general give us a huge number of sets of matrices about which we can now start asking very geometric questions, for example: how many membranes are inside the black hole?
One good reason to do this is that general consideration of black hole entropy for non-extremal black holes suggests that they are made of a gas of brane anti-brane pairs \cite{Danielsson:2001xe} and their excitations. 

\begin{figure}[ht]
\includegraphics[scale=0.3]{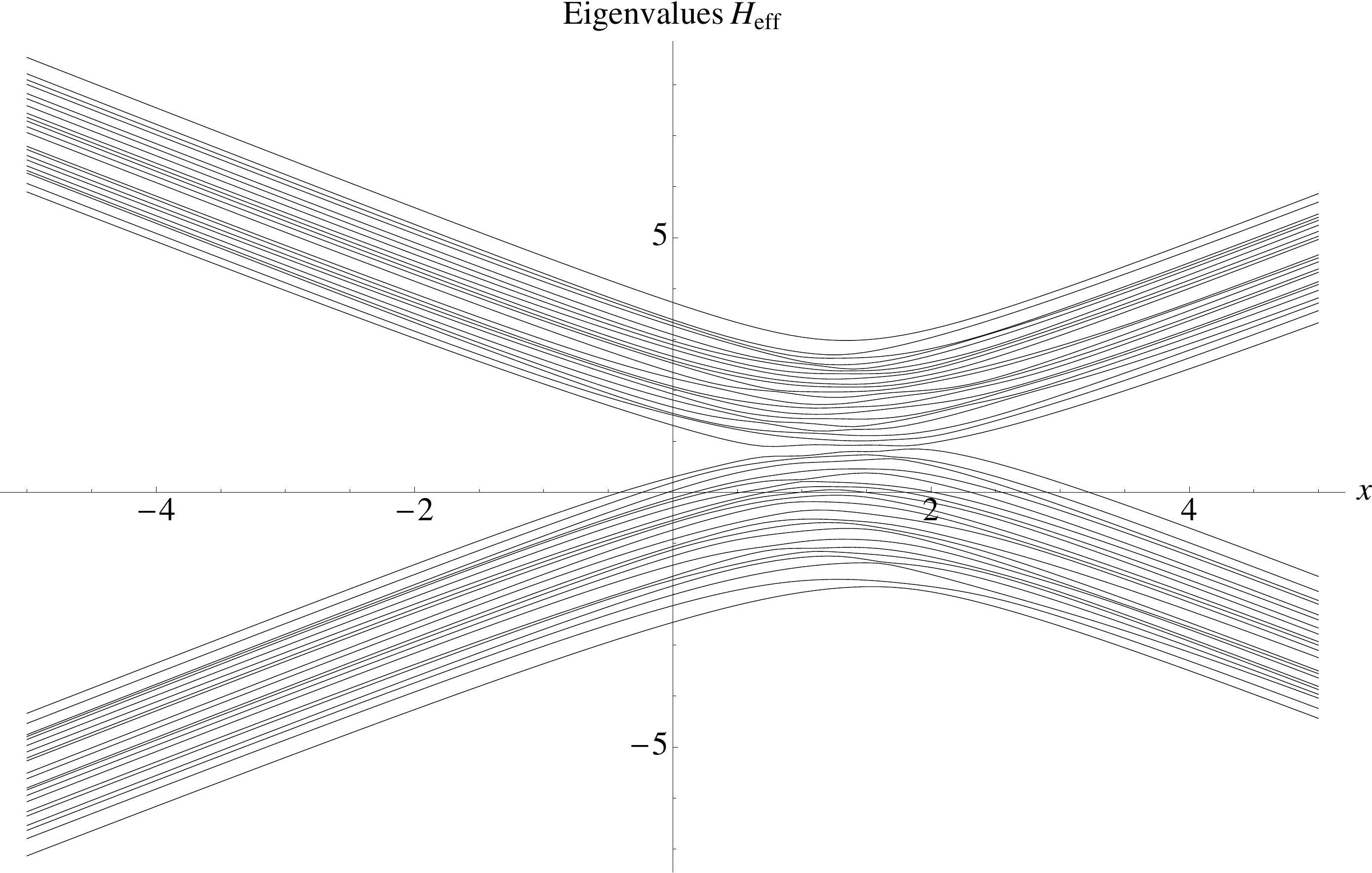}
\caption{Eigenvalues of $H_{eff}(x,0,0)_{X,Y,Z}$ from a typical configuration of matrices after thermalization, when varying $x$. The matrices have rank $21$. }
\label{fig:eigenx}
\end{figure}

Since our construction permits us to study the geometry of the typical matrices in a thermal ensemble, we can ask how does the black hole look like in the matrix variables. We will show an example of this based on data obtained from simulations similar to those reported in \cite{ABT}, where we truncate to only three matrices as described previously.  The simplest thing to do in order to understand the data is to compute the spectrum of $H_{eff}(x,y,z)_{X,Y,Z}$ given a collection of three matrices $X,Y,Z$ that are obtained from such simulations. We show the results by fixing the matrix, and setting $y,z=0$. A typical such result is shown in figure \ref{fig:eigenx}.

What we should notice is that there are various crossings of zero, and that there is a region in the center
of the configurations where the eigenvalues of $H_{eff}$ do not seem to have a gap in them. That is the black hole region of the configuration. For large $x$ we see that the eigenvalues behave as parallel lines and this matches our expectations based on perturbation theory from equation \eqref{eq:perturbationtheory}. We clearly see various crossings of zero, mostly because of the shift of flux. Also, in the region without a gap the eigenvalue distribution appears to have a well defined density of eigenvalues.  

One can show that if one varies the classical temperature of the BMN ensemble $T$ and one makes it large (so that the quartic term in the potential dominates over the cubic and quadratic terms), then the matrices themselves grow roughly like $T^{1/4}$. This is an application of the virial theorem applied to the BMN Hamiltonian. Similarly, one can show that at $T$ fixed, and varying $N$, the matrices grow only like $N^{1/4}$.  The virial theorem would just state that
\begin{equation}
E_{kin} \simeq N^2 T \simeq E_{pot}\simeq \Tr( [X,X]^2)\simeq  Tr(X^4)\simeq N x^4
\end{equation}
this is, the kinetic energy is proportional to the potential energy, which is roughly characterized by the typical eigenvalues of the matrix $X$, which we call $x$ in the equation. Since there are $N$ such eigenvalues and the potential is quartic, we expect that the result is roughly $Nx^4$. We then get $x\simeq N^{1/4}T^{1/4}$.

This means that the matrix $H_{eff}$ also grows roughly like $N^{1/4}$, so it's eigenvalues scale like $N^{1/4}$. If we assume that in the ungapped region the Hamiltonian $H_{eff}$ behaves like a random matrix in that it has a well defined density of eigenvalues when we take $N\to \infty$, then since we have $2N$ eigenvalues, the eigenvalue density near the black hole region near zero grows  like $N^{3/4}$. Because of the flux contribution to the fermion mass (the Myers effect \cite{Myers}), the center of the configuration is displaced from zero: the eigenvalue configuration is centered at $3/4$ in our units (see appendix). Far away, half the eigenvalues are above the $x$ axis, and half are below, so roughly $\rho(0) \simeq O(N^{3/4})$ eigenvalues cross zero. This means the inside of the black hole is full of branes that have been polarized, all with the same orientation. As the temperature is increased,  the eigenvalue distribution becomes wider and fewer eigenvalues cross zero. The $N$ dependence is still correct, but there is also a temperature dependence.    This polarization into D-branes is mostly because of  the Myers effect.
Remember that in this problem we have truncated to three matrices, so we are actually working on an orbifold: we are only using $H_{eff}$ with only Pauli matrices. The true fermion Hamiltonian in the BMN matrix model uses all matrices and will have different characteristics. Thus, if we truncate this way we are working with something that resembles more a brane-world black hole (we can't move it away from some locus).

\subsection{Black holes absorb matter}

An important characteristic of black holes is that if one throws matter at them, then the matter does not come out at the other side. Let us throw a fractional D0 brane at such a (orbifold) black hole. Notice that we need to do so in the orbifold of the BMN geometry. We can require the fractional  D0 brane to be at a large distance from the black hole (let us say k times the size of the black hole itself). The energy of such a D0 brane in the BMN matrix model that starts at rest is of order $k^2 N^{1/2}$. To estimate this we just look at the quadratic potential term for the $X^{1,2,3}$ matrices.

If we throw a fractional D0 brane to the black hole as described above, at each zero eigenvalue crossing a string is created due to the Hanany-Witten effect we have discussed previously. This is identical to the creation of strings observed in the one dimensional model for D-brane scattering studied in  \cite{DKPS}.
 There are about $N^{3/4}$ such strings per D0 brane that are created (as many going in as out). 
When we reach the end of the matrix configuration, these strings have a length of order $N^{1/4}$, so the energy stored in these strings is of order $\hbar N$. So long as $\hbar N >>k^2 N^{1/2}$, we find that the D0 brane does not have enough energy to come out: it gets transfered to the strings. Notice that this depends on $\hbar$. When we take $N\to \infty$, it's  clear that the strings win over the
initial energy of the D0 brane. Thus we find that the matrix thermal configuration does become a very good absorber: every fractional D0 brane that is thrown at it is eaten. Indeed, if we throw bigger objects at the thermal ensemble (let us say made of M fractional D0 branes),  the effect is proportional to the number of fractional D0 branes making the object: the linking number will guarantee that. Thus all objects are absorbed with the same efficiency. This is very similar to how black holes actually operate.

Notice also that usually in these dynamical setups if fermions are created by dynamics, then so are bosons. The accounting might be different, but they usually follow each other somewhat. Indeed, in the BMN model alone, the presence of tachyons in some regions of the dynamics can generate large numbers of bosons \cite{BT}.  Thus one should also expect bosonic modes to be created by dynamical mechanisms (the modes become non-adiabatic) rather than by a simple topological argument in general.

This simple accounting of how objects are absorbed that we found is different than other approaches that presume the formation of a tachyon in an ensemble \cite{KL}. Maybe an effective tachyon can be thought of as a collective effect of all these fermions and bosons. 

We should remind the reader that we should not take the arguments above based on generalizations of the Hanany-Witten effect very seriously for the full BMN matrix model. 
There the dynamics of the other matrices might change the physics substantially, as there we expect these D2-branes to fluctuate in the transverse directions. Thus, the topology of the Hanany-Witten effect would only be available for D8 branes, rather than D2-branes, so that there is no background flux reason to polarize D8 branes in large numbers. 

\section{Conclusion}\label{sec:conc}
The 16 supersymmetries of the BFSS matrix model can be reduced down to 4 supersymmetries, removing six of the nine bosonic  matrices and thereby giving three matrices which capture some dynamics of the full theory.
Such a reduction can be obtained by orbifolding six of the nine directions.
If the orbifold is chosen with respect to a $\BZ_k$ action, then chiral fermions arise.
Chiral fermions give rise to anomalies in four dimensions, and it follows that the fermions can encode some topological information in the reduced matrix model.
This information can be used to study the geometry of membranes formed by thermalized black holes in numerical simulations of the matrix models.
 
Appending a D0-brane probe, described by a point in $\BR^3$, to the three relevant matrices of the matrix models allows us to look at the dynamics of the theory. 
An effective Hamiltonian derived from the interaction between chiral fermions, described as fractional branes different than those we're probing, and the branes in the configuration we are probing describes how fermionic strings are created between these branes and the D0-brane.
The energy eigenvalues are proportional to the string length and thus the minimum eigenvalue can be interpreted as the minimum distance between the D0-brane and the noncommutative configuration.
If an eigenvalue equals zero, then the probe is intersecting the membrane.
Passing through the membrane changes the number of positive and negative eigenvalues by integer increments. An index function is built that captures these crossings by taking the difference of the count of positive versus negative eigenvalues.

The index inherits the symmetries of the Hamiltonian from which it is derived; it is covariant under rotations, translations, and dilatations and is gauge invariant.
The continuity of the eigenvalue functions of a matrix imply that the index function is locally constant and defines closed oriented surfaces.
It has been shown that the index is zero at infinity. The index is also additive amongst direct sums of different matrix configurations.
Finally the index has a transformation that reverses the orientation of the membranes. These properties are all true in the BFSS matrix model.
In the BMN matrix model, there is a mass contribution to the fermions caused by the dimensional reduction of the 9 dimensional gamma matrices.
This ruins the orientation reversal and scaling properties of the index function. Furthermore, the added mass changes the shapes of the membranes slightly, which is directly related to the Myers effect \cite{Myers}.

Physically we can speak of the eigenvalues of the effective Hamiltonian as representing the energies of the fermions in different regions between the membranes; positive eigenvalues corresponding to fermion creation and negative eigenvalues corresponding to fermion annihilation.
As one crosses a membrane from higher to lower index, a fermion creation operator is transformed into a fermion annihilation operator.
In the BFSS matrix model we can view this as a generalization of the Hanany-Witten effect.
In the BMN matrix model the dynamics of flux changes the results and many of the properties of the index are modified. This can be ascribed to the Myers effect \cite{Myers}.
These crossings define the surfaces in $\BR^3$. We showed configurations that correspond to both spheres and tori. We were also able to show that the surfaces carry the information of a line bundle on them with connection (which can be calculated using Berry phase arguments). This shows that the membranes we found really behave like D2-branes. Our exploration of this issue was very sketchy, so finding how to make this correspondence precise requires more work. Indeed, we would need to see if the connection we computed also includes information of the tangent bundle of the surface or not and how to separate that part from the D-brane worldvolume spin one excitations.

We were also able to generalize this index to a linking number between two such configurations.
The linking number is also interpreted in terms of the Hanany-Witen effect: it counts how many such strings are created when trying to
separate the two configurations away from each other. 

Finally we were able to show that these surfaces can be used to analyze numerical data from simulations in the BMN matrix model and in more general setups and the data show that one can make contact with conjectures about the structure of black hole interiors as made from brane-antibrane systems. We were also able to show that with the Hanany-Witten effect, the fermions created on these surfaces could be used to stop a probe D0-brane particle in a simple model. Thus it is clear that these modes can give us a handle on black hole dynamics that do not require much effort.

After solving the problem of embeddings into three dimensions, it would be interesting to understand how this same story plays out in higher dimensions. 
One of the ways in which extended objects are understood is in terms of Berry phase dynamics \cite{Berkooz:1996is} (  for a more recent discussion see
\cite{Pedder:2008je} and references therein). The Berry phase can lead to a non-trivial vector bundle structure of the fermion excitations connecting a probe to a brane.
 One can expect that if topology requires that this structure becomes degenerate at various loci, that these loci describes extended objects: again, one has to look for fermion zero modes depending on position and at least in principle it should be possible to predict that there are degenerations in some setups. However, the story might be much more complicated, as we might require to have more than one fermion zero mode simultaneously to describe  this locus. The Berry phase dynamics associated to that setup would then be non-abelian.
 It would be nice to understand this better. This might also require using extended probe branes to see the effects. The general question will then be to understand 
generalized versions of the Hamiltonian \eqref{eq:main} and the general structure of degenerations. Our construction also suggests that in these general setups there can be a similar linking number so long as one can guarantee crossings of zero of the eigenvalues of $H_{eff}$. One can show that for even dimensions (an even number of matrices) the spectrum of fermions starting from a D0 brane probe to a configuration is mirrored: for every positive eigenvalue there is a negative one. This is because one can find a matrix similar to $\gamma^5$ in four dimensions that anti-commutes with $H_{eff}$ as given by the generalization of equation \eqref{eq:main}. This suggests that ideally we should work with an odd number of matrices to make the existence of zero modes plausible for somewhat general configurations. 

A second thing that is interesting to study is how to recover the matrices given the surfaces (perhaps with additional information on them) and the rank of the matrices. 
One could also ask if the surfaces we obtained move in a way that closely resembles the membrane dynamics once we turn on the dynamics. This might be important to understand $1/N$ effects in matrix theory. 
Also, we found that in general we could reverse orientations of branes by using complex conjugation. It would be nice to understand if a brane-antibrane pair in these models generally leads to tachyons on their worldvolume and it would be interesting to analyze how tachyon condensation would progress in these setups. Also, it would be interesting to understand this issue with a probe D0-brane on top of a D2-brane: do we always get tachyons in this way?

Also, the ideas found in \cite{HL2} suggest various generalizations to different types of matrices. These ideas have applications in condensed matter physics and the 
connections we found with string theory ideas might provide interesting ways of analyzing the condensed matter systems and their dynamics. Such changes of the structure of matrices are natural when considering orientifolds. Thus our arguments should generalize to those setups.

\section*{Acknowledgements}

We would like to thank C. Asplund for discussions. D.B. would also like to thank M. Hastings and J. Maldacena for discussions. Work supported in part by DOE under grant DE-FG02-91ER40618. D.B. work supported in part by the National Science Foundation under Grant No. NSF Phy05-51164

\appendix

\section{BFSS and BMN model conventions}\label{sec:app}

The following was taken from \cite{BMN, DSJR,BT} using a mix of conventions:
\begin{align}
S &= S_0 + S_{mass} \\
S_0 &= \int dt\, \Tr\lb \sum_{j=1}^9 \frac{1}{2(2R)}(D_0\phi^j)^2 + \frac{i}{2}\Psi^\dagger D_0\Psi + \frac{(2R)}{4}\sum_{j,k=1}^9[\phi^j,\phi^k]^2 \rr \nonumber \\
&\quad \lr + \sum_{j=1}^9 \frac{1}{2}(2R)(\Psi^\dagger\gamma^j[\phi^j,\Psi]) \rb \\
S_{mass} &= \int dt\, \Tr\lb \frac{1}{2(2R)}\lp -\lb \frac{\mu}{3}\rb^2\sum_{j=1}^3(\phi^j)^2 - \lb\frac{\mu}{6}\rb^2\sum_{j=4}^9(\phi^j)^2\rp - \frac{i}{2}\lp\frac{\mu}{4}\rp\Psi^\dagger\gamma_{123}\Psi \rr \nonumber \\
&\quad \lr - \frac{\mu}{3}i\sum_{j,k,l=1}^3\epsilon_{jkl}\phi^j\phi^k\phi^l\rb
\end{align}
The fermion representation we choose to work in is not explicitly real, and so $\Psi^T$ is replaced by $\Psi^\dagger$ (see section \ref{fermiondecomposition}). Rescale to get rid of $R$:
\beq
\phi \rightarrow g^{-2/3} \phi, \quad \Psi \rightarrow \frac{1}{g}\Psi,\quad t\rightarrow\frac{g^{2/3}}{2R} t,\quad \mu\rightarrow 6Rg^{-2/3}\mu
\eeq
\begin{align*}
S &= S_0 + S_{mass} \\
S_0 &= \frac{1}{g^2}\int dt\, \Tr\lb \sum_{j=1}^9 \frac{1}{2}(D_0\phi^j)^2 + \frac{i}{2}\Psi^\dagger D_0\Psi + \frac{1}{4}\sum_{j,k=1}^9[\phi^j,\phi^k]^2 + \sum_{j=1}^9 \frac{1}{2}(\Psi^\dagger\gamma^j[\phi^j, \Psi]) \rb \\
S_{mass} &= \frac{1}{g^2}\int dt\, \Tr\lb \frac{1}{2}\lp -\mu^2\sum_{j=1}^3(\phi^j)^2 - \lb\frac{\mu}{2}\rb^2\sum_{j=4}^9(\phi^j)^2\rp - \frac{i}{2}\lp\frac{3\mu}{4}\rp\Psi^\dagger\gamma_{123}\Psi \rr \\
&\quad \lr - \mu i\sum_{j,k,l=1}^3\epsilon_{jkl}\phi^j\phi^k\phi^l\rb
\end{align*}
and again to get rid of $\mu$:
\beq
\phi \rightarrow \mu \phi, \quad \Psi \rightarrow \mu^{3/2}\Psi,\quad t\rightarrow \frac{1}{\mu} t,\quad g\rightarrow \mu^{3/2}g
\eeq
\begin{align}
S &= S_0 + S_{mass} \\
S_0 &= \frac{1}{g^2}\int dt\, \Tr\lb \sum_{j=1}^9 \frac{1}{2}(D_0\phi^j)^2 + \frac{i}{2}\Psi^\dagger D_0\Psi + \frac{1}{4}\sum_{j,k=1}^9[\phi^j,\phi^k]^2 + \sum_{j=1}^9 \frac{1}{2}(\Psi^\dagger\gamma^j[\phi^j,\Psi]) \rb \\
S_{mass} &= -\frac{1}{g^2}\int dt\, \Tr\lb \frac{1}{2}\lp \sum_{j=1}^3(\phi^j)^2 + \frac{1}{2^2}\sum_{j=4}^9(\phi^j)^2\rp + \frac{i}{2}\lp\frac{3}{4}\rp\Psi^\dagger\gamma_{123}\Psi + i\sum_{j,k,l=1}^3\epsilon_{jkl}\phi^j\phi^k\phi^l\rb
\end{align}
Notice that if we start with $\mu=0$, then in the rescaling we just modify $g$ and we find that the BFSS lagrangian has no free parameters (this is the statement that the gauge coupling in $0+1$ dimensions is dimensionful, so that there is no dimensionless coupling constant).

In the $A_t = 0$ gauge, the covariant time derivatives become ordinary time derivatives. Relabel the $\phi^j$ by $X^I$. Define $X^i = \phi^i$ for $i=1,2,3$ and $Y^a = \phi^a$ for $a=1,\dots,6$. The bosonic action takes the form
\beq
S_{B} = \frac{1}{2g^2}\int dt\, \Tr\lb (\dot{X}^i)^2 + (\dot{Y}^a)^2 - (X^i)^2 - \frac{1}{4}(Y^a)^2 - 2i\epsilon_{ijk}X^iX^jX^k - \frac{1}{2}[X^I,X^J]^2\rb
\eeq
The fermionic action becomes
\beq
S_{F} = \frac{1}{g^2}\int dt\, \Tr\lb \frac{i}{2}\Psi^\dagger\dot{\Psi} - \frac{i}{2}\lp\frac{3}{4}\rp \Psi^\dagger\gamma_{123}\Psi + \frac{1}{2}\Psi^\dagger\gamma^I[X^I, \Psi] \rb
\eeq
This is how the action is written in \cite{BT}.

\section{Fermion Decomposition}
\label{fermiondecomposition}

This section comes from appendix A of reference \cite{DSJR}. Decompose the 16 component spinor as
\begin{align}
SO(16)&\rightarrow SO(6)\otimes SO(3) \simeq SU(4) \otimes SU(2) \nonumber \\
\bf{16} & \rightarrow (\bf{4}\otimes\bf{2}) \oplus (\bar{\bf{4}} \otimes \bar{\bf{2}}) \nonumber \\
\Psi &\rightarrow \psi_{I\alpha}, \psi^{\dagger J\beta}
\end{align}
where $I$, $J$ are fundamental $SU(4)$ indices and $\alpha$, $\beta$ are fundamental $SU(2)$ indices. The spinors obey the reality condition
\beq
(\psi^\dagger)^{I\alpha} = \tilde{\psi}^{I\alpha}
\eeq
which allow us to write the spinors in the stacked form
\beq
\Psi \rightarrow \begin{pmatrix} \psi_{I\alpha} \\ \epsilon_{\alpha\beta}\psi^{\dagger I\beta} \end{pmatrix} 
\eeq
The matrices $g^a_{IJ}$ are introduced to relate the inner product of $SU(4)$ to the vector of $SO(6)$ which satisfy
\beq
g^a(g^\dagger)^b + g^b(g^\dagger)^a = 2\delta^{ab}
\eeq
The gamma matrices are then written as
\beq
\gamma^i = \begin{pmatrix} -\sigma^i\otimes 1 & 0 \\ 0 & \sigma^i\otimes I \end{pmatrix}, \quad 
\gamma^a = \begin{pmatrix} 0 & 1 \otimes g^a \\ 1\otimes (g^a)^\dagger & 0 \end{pmatrix}
\eeq
The terms in the Lagrangian then decompose as
\begin{align}
\frac{i}{2}\Psi^\dagger D_0\Psi &\rightarrow i\psi^{\dagger I\alpha}D_0\psi_{I\alpha} \\
\frac{i}{2}\Psi^\dagger \gamma_{123} \Psi &\rightarrow \psi^{\dagger I\alpha}\psi_{I\alpha} \\
\frac{1}{2}\Psi^\dagger \gamma^i [X^i, \Psi] &\rightarrow -\psi^{\dagger I\alpha}\sigma^{i\beta}_{\alpha}[X^i, \psi_{I\beta}] \\
\frac{1}{2}\Psi^\dagger \gamma^a [X^a, \Psi] &\rightarrow \frac{1}{2}\epsilon_{\alpha\beta}\psi^{\dagger I\alpha}g^a_{IJ}[Y^a, \psi^{\dagger J\beta}] - \frac{1}{2}\epsilon^{\alpha\beta}\psi_{I\alpha}(g^\dagger)^{aIJ}[Y^a, \psi_{J\beta}]
\end{align}
The fermionic part of the action (in the $A_0 = 0$ gauge) may then be written as
\begin{align}
S_F &= \frac{1}{g^2}\int dt\,\Tr\lb i\psi^{\dagger I\alpha}\dot{\psi}_{I\alpha} - \frac{3}{4}\psi^{\dagger I\alpha}\psi_{I\alpha} -\psi^{\dagger I\alpha}\sigma^{i\beta}_{\alpha}[X^i, \psi_{I\beta}] \rr \nonumber \\
&\qquad \lr + \frac{1}{2}\epsilon_{\alpha\beta}\psi^{\dagger I\alpha}g^a_{IJ}[Y^a, \psi^{\dagger J\beta}] - \frac{1}{2}\epsilon^{\alpha\beta}\psi_{I\alpha}(g^\dagger)^{aIJ}[Y^a, \psi_{J\beta}] \rb
\end{align}

Notice that the coupling to the $X$ variables uses just the Pauli matrices after this decomposition. Also, a 
$\psi$ spinor is always paired with its conjugate. If we perform orbifolds that are chiral, this structure remains, but the  other mass terms that do not preserve four dimensional chirality might be eliminated.

\section{Fermionic Modes between displaced fuzzy spheres}

For our paper we want to consider computing fermionic modes between two fuzzy spheres in the BMN matrix model that have been displaced as described in \cite{BT}. We want to restrict to a chiral projection of the modes between two such fuzzy spheres. First we will setup some conventions for the fermionic modes of a single fuzzy sphere. Then we work with the more general problem.

\subsection{Diagonal Fermionic modes}

This section is essentially a repeat of section 5.2 of \cite{DSJR} using the conventions of \cite{BT}. 

The $SU(4)$ indices are dropped as they do not come into play at all during the following calculation. We take the following conventions for the spherical harmonics and the angular momentum generators:
\begin{align*}
\Tr(Y^\dagger_{lm}Y_{l'm'}) &= \frac{1}{2}\delta_{ll'}\delta_{mm'} \\
\Lambda_+^{lm} &= \sqrt{(l + m)(l - m + 1)} \\
\Lambda_-^{lm} &= \sqrt{(l - m)(l + m + 1)}
\end{align*}
\vspace{-0.60in}
\begin{align*}
[L^3, Y_{lm}] &= mY_{lm} & [L^3, Y_{lm}^\dagger] &= -mY^\dagger_{lm} \\
[L^+, Y_{lm}] &= \Lambda_-^{lm}Y_{lm+1} & [L^+, Y_{lm}^\dagger] &= -\Lambda_+^{lm}Y_{lm-1}^\dagger \\
[L^-, Y_{lm}] &= \Lambda_+^{lm}Y_{lm-1} & [L^-, Y_{lm}^\dagger] &= -\Lambda_-^{lm}Y_{lm+1}^\dagger
\end{align*}
\vspace{-0.35in}
\beqs
\Lambda_+^{l-l} = 0 \quad \Lambda_-^{ll} = 0 \quad  \Lambda_+^{lm+1} = \Lambda_-^{lm}
\eeqs
where $L^{\pm} = L_1 \pm iL_2$. We expand the fermions as
\beq
\psi_\alpha = \sum_{lm} \psi_\alpha^{lm}Y_{lm}
\eeq
The potential in the presence of the bosonic VEV's becomes
\begin{align*}
V_F - \frac{3}{4}\Tr(\psi^{\dagger\alpha}\psi_{\alpha}) &= \Tr\lb \psi^{\dagger\alpha} \sigma^{i\beta}_\alpha[L^i,\psi_{I\beta}]\rb \\
&= \Tr\lb \psi^{\dagger +}\lp [L^3, \psi_+] + [L^-, \psi_-] \rp + \psi^{\dagger -}\lp [L^+, \psi_+] - [L^3, \psi_-] \rp \rb \\
&= \Tr\lb \psi^{\dagger +}\sum_{lm} \lp m\psi_+^{lm}Y_{lm} + \Lambda_+^{lm} \psi_-^{lm}Y_{lm-1}\rp + \psi^{\dagger -}\sum_{lm} \lp \Lambda_-^{lm} \psi_+^{lm}Y_{lm+1} - m\psi_-^{lm}Y_{lm}\rp \rb \\
&= \frac{1}{2}\sum_{lmm'}\lb \psi_+^{\dagger lm'}\lp m\psi_+^{lm}\delta_{m'm} + \Lambda_+^{lm} \psi_-^{lm}\delta_{m'm-1}\rp + \psi_-^{\dagger lm'}\lp \Lambda_-^{lm} \psi_+^{lm}\delta_{m'm+1} - m\psi_-^{lm}\delta_{m'm}\rp \rb \\
&= \frac{1}{2}\sum_{lmm'} \begin{pmatrix} \psi_+^{\dagger lm'} & \psi_-^{\dagger lm'} \end{pmatrix}
\begin{pmatrix}
m & \Lambda_+^{lm} \delta_{m'm-1} \\
\Lambda_-^{lm} \delta_{m'm+1} & -m
\end{pmatrix}
\begin{pmatrix} \psi_+^{lm} \\ \psi_-^{lm} \end{pmatrix}
\end{align*}
The eigenvalues of the matrix plus $3/4$ give the mass spectrum. Note that $-l\leq m,m'\leq l$. The $\delta$'s tell us that this matrix has $2l$ two by two blocks and two one by one blocks where $m,m' = l$ and $m,m' = -l$. Each of the one by one blocks yield the eigenvalue $l$. The $2l$ two by two blocks can be parametrized according to $m$ from $-l$ to $l - 1$. They are given by
\beq
\begin{pmatrix}
m & \Lambda_+^{lm+1} \\
\Lambda_-^{lm} & -(m + 1)
\end{pmatrix} = 
\begin{pmatrix}
m & \Lambda_-^{lm} \\
\Lambda_-^{lm} & -(m + 1)
\end{pmatrix}
\eeq
The eigenvalues are given by the characteristic equation
\begin{align*}
0 &= (m - \lambda)(-m - 1 - \lambda) - (l - m)(l + m + 1) = (\lambda - l)(\lambda + l + 1)
\end{align*}
Thus the eigenvalues are $l$ and $-(l+1)$. This means that the mass spectrum is $M = 3/4 + l$ with degeneracy of $2l + 2$ and $M = -(l + 1/4)$ with degeneracy $2l$. Note that there are two more positive eigenvalues than negative eigenvalues.

\subsection{Off-diagonal modes}

Here we follow the procedure of the previous section and that in \cite{BT}. Expand the off diagonal modes in fuzzy monopole harmonics
\beq
\psi_{\alpha} = \sum_{lm}
\begin{pmatrix}
0 & \delta\psi_{\alpha}^{lm} Y_{lm} \\
(\delta\tilde\psi_{\alpha}^{lm}) Y_{lm}^\dagger & 0 
\end{pmatrix}
\eeq
The following commutators are necessary
\begin{align*}
[X^3, \psi_\alpha] &= 
\sum_{lm}
\begin{pmatrix}
0 & \delta\psi_{\alpha}^{lm}[L^3, Y_{lm}] \\
(\delta\tilde\psi_{\alpha}^{lm})[L^3, Y^\dagger_{lm}] & 0
\end{pmatrix} + b \lb\begin{pmatrix}
0 & 0 \\
0 & 1
\end{pmatrix}, \begin{pmatrix}
0 & \delta\psi_{\alpha}^{lm} Y_{lm} \\
(\delta\tilde\psi_{\alpha}^{lm}) Y^\dagger_{lm} & 0
\end{pmatrix}\rb \\
&= \sum_{lm}
\begin{pmatrix}
0 & (m-b)\delta\psi_{\alpha}^{lm} Y_{lm} \\
-(m-b)(\delta\tilde\psi_{\alpha}^{lm})Y_{lm}^\dagger & 0 
\end{pmatrix} \\
[X^+, \psi_\alpha] &= 
\sum_{lm}
\begin{pmatrix}
0 & \delta\psi_{\alpha}^{lm}[L^+, Y_{lm}] \\
(\delta\tilde\psi_{\alpha}^{lm})[L^+, Y^\dagger_{lm}] & 0
\end{pmatrix} 
= \sum_{lm}
\begin{pmatrix}
0 & \Lambda^{lm}_-\delta\psi_{\alpha}^{lm} Y_{lm+1} \\
-\Lambda^{lm}_+(\delta\tilde\psi_{\alpha}^{lm}) Y_{lm-1}^\dagger & 0 
\end{pmatrix} \\
[X^-, \psi_\alpha] &= 
\sum_{lm}
\begin{pmatrix}
0 & \delta\psi_{\alpha}^{lm}[L^-, Y_{lm}] \\
(\delta\tilde\psi_{\alpha}^{lm})[L^-, Y^\dagger_{lm}] & 0
\end{pmatrix} 
= \sum_{lm}
\begin{pmatrix}
0 & \Lambda_+^{lm}\delta\psi_{\alpha}^{lm} Y_{lm-1} \\
-\Lambda_-^{lm}(\delta\tilde\psi_{\alpha}^{lm})Y_{lm+1}^\dagger & 0 
\end{pmatrix}
\end{align*}
Substituting these expressions into the potential, taking the chiral projection and finally taking the trace we have
\begin{align}
V_F - \frac{3}{4}\Tr(\psi^{\dagger\alpha}\psi_{\alpha})= \begin{pmatrix} \psi^{\dagger+ lm'} & \psi^{\dagger- lm'} \end{pmatrix}
\begin{pmatrix}
m -b & \Lambda_+^{lm} \delta_{m'm-1} \\
\Lambda_-^{lm} \delta_{m'm+1} & -(m - b)
\end{pmatrix}
\begin{pmatrix} \psi_+^{lm} \\ \psi_-^{lm} \end{pmatrix}
\end{align}
Notice that this essentially produces the exact same matrix system as for the diagonal modes except with different diagonal elements. Also, half spin objects are allowed as when we decompose into the tensor product we can get different spins. There is a one by one block with $\lambda = l - b$, another with $\lambda = l + b$, and $2l$ two by two blocks. The matrix for these blocks is
\beq
\begin{pmatrix}
m -b & \Lambda_-^{lm}  \\
\Lambda_-^{lm} & -(m + 1 - b)
\end{pmatrix}
\eeq
with $-l\leq m \leq l - 1$. The eigenvalues satisfy
\begin{align*}
0 &= -(m - b - \lambda)(m + 1 - b + \lambda) - (l - m)(l + m - 1) \\
&= \lambda^2 + \lambda - l(l+1) - b(b-1) + 2mb
\end{align*}
Solving for $\lambda$ gives
\begin{align}
\lambda = -\frac{1}{2}\pm \sqrt{(l - m)(l + m + 1) + (b - m - 1/2)^2}
\end{align}
Thus the full modes are with $-l\leq m\leq l-1$ (each with degeneracy two):
\mybox{m = \frac{1}{4}\pm\sqrt{(l - m)(l + m + 1) + (b - m - 1/2)^2}}
We also have two other modes corresponding to the one by one blocks of the mass matrix: $m = 3/4 + l \pm b$. These yield zero modes for the right value of $b$. That is we have zero modes when
\mybox{b = \pm(l + 3/4)}
for the modes with the greatest angular momentum in the z direction for a given value of $\ell$. These zero modes are correlated with the modes that become tachyonic for bosons in the same type of configurations found in \cite{BT}: they are objects of maximal spin fixing $\ell$.


\begin{thebibliography}{99}

\bibitem{DLP} 
  J.~Dai, R.~G.~Leigh and J.~Polchinski,
  ``New Connections Between String Theories,''
  Mod.\ Phys.\ Lett.\ A {\bf 4}, 2073 (1989).
  J.~Polchinski,
  ``Dirichlet Branes and Ramond-Ramond charges,''
  Phys.\ Rev.\ Lett.\  {\bf 75}, 4724 (1995)
  [hep-th/9510017].


\bibitem{Hoppe}
J. ~ Hoppe, ``Quantum theory of a massless relativistic surface and a two-dimensional bound state problem'' MIT phD Thesis, 1982.

\bibitem{de Wit:1988ig} 
  B.~de Wit, J.~Hoppe and H.~Nicolai,
  ``On the Quantum Mechanics of Supermembranes,''
  Nucl.\ Phys.\ B {\bf 305}, 545 (1988).


\bibitem{BFSS} 
  T.~Banks, W.~Fischler, S.~H.~Shenker and L.~Susskind,
  ``M theory as a matrix model: A Conjecture,''
  Phys.\ Rev.\ D {\bf 55}, 5112 (1997)
  [hep-th/9610043].


\bibitem{ABT} 
  C.~Asplund, D.~Berenstein and D.~Trancanelli,
  ``Evidence for fast thermalization in the BMN matrix model,''
  Phys.\ Rev.\ Lett.\  {\bf 107}, 171602 (2011)
  [arXiv:1104.5469 [hep-th]].


\bibitem{AHHS} 
  T.~Azeyanagi, M.~Hanada, T.~Hirata and H.~Shimada,
  ``On the shape of a D-brane bound state and its topology change,''
  JHEP {\bf 0903}, 121 (2009)
  [arXiv:0901.4073 [hep-th]].




\bibitem{BMN} 
  D.~E.~Berenstein, J.~M.~Maldacena and H.~S.~Nastase,
 ``Strings in flat space and pp waves from N=4 superYang-Mills,''
  JHEP {\bf 0204}, 013 (2002)
  [hep-th/0202021].



\bibitem{HW} 
  A.~Hanany and E.~Witten,
  ``Type IIB superstrings, BPS monopoles, and three-dimensional gauge dynamics,''
  Nucl.\ Phys.\ B {\bf 492}, 152 (1997)
  [hep-th/9611230].
  
\bibitem{BDG} 
  C.~P.~Bachas, M.~R.~Douglas and M.~B.~Green,
  ``Anomalous creation of branes,''
  JHEP {\bf 9707}, 002 (1997)
  [hep-th/9705074].
  
\bibitem{Banks:1996nn} 
  T.~Banks, N.~Seiberg and S.~H.~Shenker,
  ``Branes from matrices,''
  Nucl.\ Phys.\ B {\bf 490}, 91 (1997)
  [hep-th/9612157].
 
 
\bibitem{Steinacker:2010rh} 
  H.~Steinacker,
  ``Emergent Geometry and Gravity from Matrix Models: an Introduction,''
  Class.\ Quant.\ Grav.\  {\bf 27}, 133001 (2010)
  [arXiv:1003.4134 [hep-th]].
  
  
\bibitem{Taylor} 
  D.~N.~Kabat and W.~Taylor,
  ``Linearized supergravity from matrix theory,''
  Phys.\ Lett.\ B {\bf 426}, 297 (1998)
  [hep-th/9712185].
  W.~Taylor and M.~Van Raamsdonk,
  ``Multiple D0-branes in weakly curved backgrounds,''
  Nucl.\ Phys.\ B {\bf 558}, 63 (1999)
  [hep-th/9904095].
  K.~Millar, W.~Taylor and M.~Van Raamsdonk,
  ``D particle polarizations with multipole moments of higher dimensional branes,''
  hep-th/0007157.
  

  
\bibitem{Taylorreview} 
  W.~Taylor,
  ``M(atrix) theory: Matrix quantum mechanics as a fundamental theory,''
  Rev.\ Mod.\ Phys.\  {\bf 73}, 419 (2001)
  [hep-th/0101126].
  
  
  
\bibitem{Kabat:1997im} 
  D.~N.~Kabat and W.~Taylor,
  ``Spherical membranes in matrix theory,''
  Adv.\ Theor.\ Math.\ Phys.\  {\bf 2}, 181 (1998)
  [hep-th/9711078].
  
\bibitem{DM} 
  M.~R.~Douglas and G.~W.~Moore,
  ``D-branes, quivers, and ALE instantons,''
  hep-th/9603167.
  
\bibitem{KaS} 
  S.~Kachru and E.~Silverstein,
  ``4-D conformal theories and strings on orbifolds,''
  Phys.\ Rev.\ Lett.\  {\bf 80}, 4855 (1998)
  [hep-th/9802183].
  
\bibitem{DGM} 
  M.~R.~Douglas, B.~R.~Greene and D.~R.~Morrison,
  ``Orbifold resolution by D-branes,''
  Nucl.\ Phys.\ B {\bf 506}, 84 (1997)
  [hep-th/9704151].


\bibitem{BL} 
  D.~Berenstein and R.~G.~Leigh,
  ``Resolution of stringy singularities by noncommutative algebras,''
  JHEP {\bf 0106}, 030 (2001)
  [hep-th/0105229].

\bibitem{KKP} 
  N.~Kim, T.~Klose and J.~Plefka,
  ``Plane wave matrix theory from N=4 superYang-Mills on R x S**3,''
  Nucl.\ Phys.\ B {\bf 671}, 359 (2003)
  [hep-th/0306054].

\bibitem{Myers} 
  R.~C.~Myers,
  ``Dielectric branes,''
  JHEP {\bf 9912}, 022 (1999)
  [hep-th/9910053].

  
 \bibitem{HL} 
  M.~B.~Hastings and T.~A.~Loring
  ``Almost commuting matrices, localized {W}annier functions, and
              the quantum {H}all effect,"
 J. Math. Phys. {\bf 51}, 2010, 015214
 [arxiv:0910.5490]

\bibitem{HL2}
  M.~B.~Hastings and T.~A.~Loring.
  ``Topological insulators and $C^*$ algebras: Theory and Numerical Practice"
Ann. Physics {\bf 326} (2011) 1699-1759  
	[arXiv:1012.1019]


\bibitem{DelgadilloBlando:2012xg} 
  R.~Delgadillo-Blando and D.~O'Connor,
  ``Matrix geometries and Matrix Models,''
  arXiv:1203.6901 [hep-th].


\bibitem{Steinacker:2003sd} 
  H.~Steinacker,
  ``Quantized gauge theory on the fuzzy sphere as random matrix model,''
  Nucl.\ Phys.\ B {\bf 679}, 66 (2004)
  [hep-th/0307075].


\bibitem{BJL} 
  D.~Berenstein, V.~Jejjala and R.~G.~Leigh,
  ``Noncommutative moduli spaces, dielectric tori and T duality,''
  Phys.\ Lett.\ B {\bf 493}, 162 (2000)
  [hep-th/0006168].

\bibitem{NT} 
  T.~Nishioka and T.~Takayanagi,
 ``Fuzzy Ring from M2-brane Giant Torus,''
  JHEP {\bf 0810}, 082 (2008)
  [arXiv:0808.2691 [hep-th]].

\bibitem{Arnlind:2006ux} 
  J.~Arnlind, M.~Bordemann, L.~Hofer, J.~Hoppe and H.~Shimada,
  ``Fuzzy Riemann surfaces,''
  JHEP {\bf 0906}, 047 (2009)
  [hep-th/0602290].
  
  
  \bibitem{Shimadaft}
J.~Arnlind, M.~ Bordemann, L.~Hofer, J.~Hoppe, H.~Shimada,
``Noncommutative Riemann Surfaces"
  arXiv:0711.2588v1 [math-ph]
  
\bibitem{DSJR} 
  K.~Dasgupta, M.~M.~Sheikh-Jabbari and M.~Van Raamsdonk,
  ``Matrix perturbation theory for M theory on a PP wave,''
  JHEP {\bf 0205}, 056 (2002)
  [hep-th/0205185].
  
  
  
  
\bibitem{Shimada:2003ks} 
  H.~Shimada,
  ``Membrane topology and matrix regularization,''
  Nucl.\ Phys.\ B {\bf 685}, 297 (2004)
  [hep-th/0307058].
  
  
\bibitem{Shimada:2008xy} 
  H.~Shimada,
  ``beta-deformation for matrix model of M-theory,''
  Nucl.\ Phys.\ B {\bf 813}, 283 (2009)
  [arXiv:0804.3236 [hep-th]].

\bibitem{AHH} 
  J.~Arnlind, J.~Hoppe and G.~Huisken,
  ``Discrete curvature and the Gauss-Bonnet theorem,''
  arXiv:1001.2223 [math-ph].



\bibitem{TR} 
  W.~Taylor and M.~Van Raamsdonk,
  ``Supergravity currents and linearized interactions for matrix theory configurations with fermionic backgrounds,''
  JHEP {\bf 9904}, 013 (1999)
  [hep-th/9812239].



\bibitem{Chatzistavrakidis:2011gs} 
  A.~Chatzistavrakidis, H.~Steinacker and G.~Zoupanos,
  ``Intersecting branes and a standard model realization in matrix models,''
  JHEP {\bf 1109}, 115 (2011)
  [arXiv:1107.0265 [hep-th]].

\bibitem{Berkooz:1996km} 
  M.~Berkooz, M.~R.~Douglas and R.~G.~Leigh,
  ``Branes intersecting at angles,''
  Nucl.\ Phys.\ B {\bf 480}, 265 (1996)
  [hep-th/9606139].

\bibitem{IMSY} 
  N.~Itzhaki, J.~M.~Maldacena, J.~Sonnenschein and S.~Yankielowicz,
  ``Supergravity and the large N limit of theories with sixteen supercharges,''
  Phys.\ Rev.\ D {\bf 58}, 046004 (1998)
  [hep-th/9802042 [math-ph]].


\bibitem{CW} 
  S.~Catterall and T.~Wiseman,
  ``Towards lattice simulation of the gauge theory duals to black holes and hot strings,''
  JHEP {\bf 0712}, 104 (2007)
  [arXiv:0706.3518 [hep-lat]].

\bibitem{AHNT} 
  K.~N.~Anagnostopoulos, M.~Hanada, J.~Nishimura and S.~Takeuchi,
  ``Monte Carlo studies of supersymmetric matrix quantum mechanics with sixteen supercharges at finite temperature,''
  Phys.\ Rev.\ Lett.\  {\bf 100}, 021601 (2008)
  [arXiv:0707.4454 [hep-th]].


\bibitem{HHNT} 
  M.~Hanada, Y.~Hyakutake, J.~Nishimura and S.~Takeuchi,
  ``Higher derivative corrections to black hole thermodynamics from supersymmetric matrix quantum mechanics,''
  Phys.\ Rev.\ Lett.\  {\bf 102}, 191602 (2009)
  [arXiv:0811.3102 [hep-th]].

\bibitem{CVa} 
  S.~Catterall and G.~van Anders,
  ``First Results from Lattice Simulation of the PWMM,''
  JHEP {\bf 1009}, 088 (2010)
  [arXiv:1003.4952 [hep-th]].



\bibitem{DKPS} 
  M.~R.~Douglas, D.~N.~Kabat, P.~Pouliot and S.~H.~Shenker,
  ``D-branes and short distances in string theory,''
  Nucl.\ Phys.\ B {\bf 485}, 85 (1997)
  [hep-th/9608024].
 
\bibitem{Danielsson:2001xe} 
  U.~H.~Danielsson, A.~Guijosa and M.~Kruczenski,
  ``Brane anti-brane systems at finite temperature and the entropy of black branes,''
  JHEP {\bf 0109}, 011 (2001)
  [hep-th/0106201].

\bibitem{BT} 
  D.~Berenstein and D.~Trancanelli,
  ``Dynamical tachyons on fuzzy spheres,''
  Phys.\ Rev.\ D {\bf 83}, 106001 (2011)
  [arXiv:1011.2749 [hep-th]].




\bibitem{KL} 
  D.~N.~Kabat and G.~Lifschytz,
  ``Tachyons and black hole horizons in gauge theory,''
  JHEP {\bf 9812}, 002 (1998)
  [hep-th/9806214].







\bibitem{Berkooz:1996is} 
  M.~Berkooz and M.~R.~Douglas,
  ``Five-branes in M(atrix) theory,''
  Phys.\ Lett.\ B {\bf 395}, 196 (1997)
  [hep-th/9610236].


\bibitem{Pedder:2008je} 
  C.~Pedder, J.~Sonner and D.~Tong,
  ``The Berry Phase of D0-Branes,''
  JHEP {\bf 0803}, 065 (2008)
  [arXiv:0801.1813 [hep-th]].



  

  

  

  

  


 
 
  
 

 

\end{thebibliography}
\end{document}